\documentclass[11pt, a4paper]{article}

\usepackage{graphicx}    
\usepackage{epsfig}
\usepackage{amssymb,amsmath}
\usepackage{setspace,subfigure}
\usepackage{color}

\newcommand{\gerda}    {{\sc Gerda }}

\begin{document}

\title{Status of the GERDA experiment aimed to search for neutrinoless double beta decay of $^{76}$Ge}

\author{Anatoly A. Smolnikov for the GERDA collaboration}

\date{}

\maketitle


\begin{abstract}

The progress in the development of the new international \gerda (GErmanium Detector Array) experiment is presented. Main purpose of the experiment is to search for the neutrinoless double beta decay of $^{76}$Ge. The experimental set up is under construction in the underground laboratory of LNGS. \gerda will operate with bare germanium semiconductor detectors (enriched in $^{76}$Ge) situated in liquid argon. In the Phase I the existing enriched detectors from the previous Heidelberg-Moscow and IGEX experiments are employed, in the Phase II the new segmented detectors made from recently produced enriched material will be added. Novel concepts for background suppression including detector segmentation and anti-coincidence with LAr scintillation are developed.

\end{abstract}

\section{Introduction}

The GERmanium Detector Array, \gerda \cite{gerda}, is designed to search for neutrinoless double beta (0$\nu\beta\beta$) decay of $^{76}$Ge. The importance of such a search is emphasized by the evidence of a non-zero neutrino mass from neutrino flavour oscillations.  The (0$\nu\beta\beta$) decay is one of the most sensitive probes of still unknown neutrino properties such as neutrino type and their mass scale. Neutrino oscillations provide only values for $\Delta m^2$, the differences of the squared masses of the neutrino mass eigenstates. Neutrinoless double beta decay can yield information on the absolute neutrino mass scale. Moreover, the neutrino (of a given flavour) could be the anti-particle of itself. Fermions of this type are called Majorana particles, in contrast to Dirac particles. The (0$\nu\beta\beta$) decay violates the global lepton number by two units ($\Delta L$=2) and it is possible only if neutrinos are Majorana particles. In the Standard Model neutrinos are mass less, however, the implications of massive neutrinos for models beyond the Standard Model differ for Majorana and Dirac neutrinos. It means that an observation of the (0$\nu\beta\beta$) decay gives an unique tool to penetrate beyond the Standard Model.

There are three main modes of $\beta\beta$ decay under consideration, namely,\\
2$\nu\beta\beta$:  (A,Z) $\rightarrow$ (A,Z + 2) + 2e$^-$ + 2$\nu$; \\
(0$\nu\beta\beta$):  (A,Z) $\rightarrow$ (A,Z + 2) + 2e$^-$; \\
0$\nu\chi^0\beta\beta$:  (A,Z) $\rightarrow$ (A,Z + 2) + 2e$^-$ + $\chi^0$ (or n$\chi^0$). \\

The two neutrino double beta decay (2$\nu\beta\beta$) is a second-order process, which is not forbidden by any conservation law. The ($2\nu\beta\beta$) decay has been detected for 10 isotopes ($^{48}$Ca, $^{76}$Ge, $^{82}$Se, $^{96}$Zr, $^{100}$Mo, $^{116}$Cd, $^{128}$Te, $^{130}$Te, $^{150}$Nd, $^{238}$U). The half-lives for this process range from T$_{1/2}=(7.7 \pm 0.7{\rm (stat)}\pm 0.8{\rm (syst)})\cdot 10^{18}$y for $^{150}$Nd \cite{Nemo} to T$_{1/2}=(7.2 \pm 0.3)\cdot 10^{24}$y for $^{128}$Te \cite{Te128}. The half life of the $2\nu\beta\beta$ decay of $^{76}$Ge is T$_{1/2}=(1.3\pm0.1)\cdot 10^{21}$~y \cite{AnnRev02}.

The neutrinoless double beta decay ($0\nu\beta\beta$) can occur through different processes but all of them require that the neutrino has nonzero mass and is a Majorana particle. The most proximate theoretical model is to mediate neutrinoless double beta decay by the exchange of a light Majorana neutrino or an admixture of right-handed currents in weak interaction. This process can be also initiated by the exchange of supersymmetric particles and is considered in some supersymmetric models.
The ($0\nu\beta\beta$) decay is a very sensitive probe for the effective Majorana neutrino mass, which is defined as $m_{\beta\beta}=\sum_i U^2_{ei}m_i$, because for the exchange of a light Majorana neutrino, the half-life for the neutrino-less double beta decay is inversely proportional to the square of the effective Majorana mass:
$$\frac{1}{T^{0\nu}_{1/2}(A,Z)}=|m_{\beta\beta}|^2|M^{0\nu}(A,Z)|^2G^{0\nu}(E_0,Z)$$
Here $M^{0\nu}(A,Z)$ is the nuclear matrix element (NME) and $G^{0\nu}(E_0,Z)$ is the known, energy dependent phase space integral, depending on the energy release $E_0$. The NME only depends on the properties of the nucleus. However, depending on the model used, there are different predictions for the value of the NME.

In the (0$\nu\chi^0\beta\beta$) decay a Majoron particle (massless Goldstone boson) can escape together with two electrons. Several other (0$\nu\chi^0\beta\beta$) models with emission of two or even more Majorons were proposed too.
     The (A, Z) - (A,Z - 2) processes also possible, for instance,  the emission of two positrons - the double $\beta^+$-decay ($2\nu\beta^+\beta^+$) and the double electron capture ($2\nu ECEC$).

The experimental signature for (0$\nu\beta\beta$) decay is the observation of a peak at the endpoint $Q_{\beta\beta}$ in the two electron sum energy spectrum. The most sensitive 0$\nu\beta\beta$-experiments so far are based on High-Purity-Germanium (HPGe) detector technology. This is due to the combination of a good energy resolution of the detectors at the Q-value of $^{76}$Ge, high purity of the Ge crystals (very low intrinsic background) and the high signal detection efficiency. The (0$\nu\beta\beta$) decay of  $^{76}$Ge shows up in the measured energy spectrum through a line at 2039 keV, the Q-value of the decay.  Semiconductor detectors fabricated from high purity Ge material enriched in $^{76}$Ge are outstanding (0$\nu\beta\beta$) decay detectors being simultaneously the $\beta\beta$ decay source and a 4$\pi$ detector with the excellent energy resolution of a few keV at Q = 2039~keV.

The best limits for (0$\nu\beta\beta$) decay in $^{76}$Ge are due to the Heidelberg-Moscow (HdM) \cite{hdhk} and IGEX \cite{igex} enriched $^{76}$Ge experiments yielding lower half-life limits of about T$_{1/2} > 1.9\cdot 10^{25}$ y and T$_{1/2} > 1.6\cdot 10^{25}$ y, correspondently, that leads to the upper limit on effective Majorana mass of $|m_{\beta\beta}| <$ 0.33 - 1.35 eV . The uncertainty in the mass limit reflects the limited knowledge on the nuclear matrix elements. A part of the HdM collaboration (KKGH) has claimed observation of (0$\nu\beta\beta$) decay in $^{76}$Ge  from a re-analysis of the obtained data, with a half-life of T$_{1/2} = 1.2^{+3.0}_{-0.5}\cdot 10^{25}$ y (3$\sigma$ range), implying a $|m_{\beta\beta}|$ value between 0.1 and 0.9 eV with the central value of 0.44 eV \cite{claim}.

It is evident that this claim needs verification by other experiments with increased sensitivity. The largest currently running experiments are CUORICINO and NEMO-3. CUORICINO located in LNGS (Italy) and uses $^{130}$Te as the double beta nucleus. The CUORICINO setup is an array of cryogenic bolometers made from Tellurite crystals with a total mass of 41 kg (33.8\% $^{130}$Te). The NEMO-3 set up (located in LSM, France) is a cylindrical detector with source foils sandwiched by tracking Geiger detectors and surrounded by a plastic scintillator calorimeter in a 25 Gauss magnetic field. The main isotopes in NEMO-3 are $^{100}$Mo (7kg) and $^{82}$Se (1kg). The sensitivities of both experiments  in terms of $|m_{\beta\beta}|$ are in the 0.5 eV range. The next stages of these experiments will be CUORE with a planned total mass of 760 kg and the Super-NEMO detector, currently conceived to contain of about 100 kg $^{150}$Nd or $^{82}$Se.  The ongoing CUORICINO and NEMO-3 experiments could confirm the (0$\nu\beta\beta$) decay signal for $^{130}$Te and $^{100}$Mo but cannot refute the KKGH claim in case of a null result due to the uncertainties of the nuclear matrix elements.

The \gerda experiment aims at probing (0$\nu\beta\beta$) decay of $^{76}$Ge with a sensitivity of T$_{1/2} > 1.5\cdot 10^{26}$ y at 90\% confidence level corresponding to a $|m_{\beta\beta}|$ range from 0.1 to 0.3 eV. Using in its first phase the refurbished $^{76}$Ge detectors of the previous HdM and IGEX experiments, a total of about 18 kg, \gerda will be able to scrutinize the recent claim for the (0$\nu\beta\beta$) decay observation with high statistical significance even after one year of running with 15~kg$\cdot$y of statistics at a background level of $\leq$10$^{-2}$~counts/(keV$\cdot$kg$\cdot$y) and to go to higher sensitivity with 150~kg$\cdot$y of statistics in the second phase at a background level of $\leq$10$^{-3}$~counts/(keV$\cdot$kg$\cdot$y) at the Q$_{\beta\beta}$ value of 2039~keV.

The main design feature of \gerda is to use the cryogenic liquid Argon (LAr) as shielding against gamma radiation \cite{heu}, the dominant background in earlier experiments. High purity germanium detectors (enriched in $^{76}$Ge) are immersed directly in the cryogenic liquid which also acts as the cooling medium. To reach the background level required for the second phase  new methods are required to suppress the intrinsic background of the detectors which mostly due to cosmogenically produced isotopes. The segmentation of the detectors, as well as pulse shape analysis and anticoincidence between nearby detectors assembled in several strings will help to identify and select background events due to multiple Compton scattering. In addition the novel concept to use the LAr scintillation light as anti-coincidence signal for further background suppression is under development.
    The \gerda setup will be installed in the Hall A of the National Gran Sasso Laboratory (LNGS) of INFN Italy (1400 m rock overburden). The \gerda collaboration consists of about 80 physicists from 13 institutions coming from 5 countries \cite{collaboration}.

\section{GERDA design, main features and expected sensitivity}

\begin{figure}[h]
\begin{center}
\includegraphics[width=0.65\linewidth]{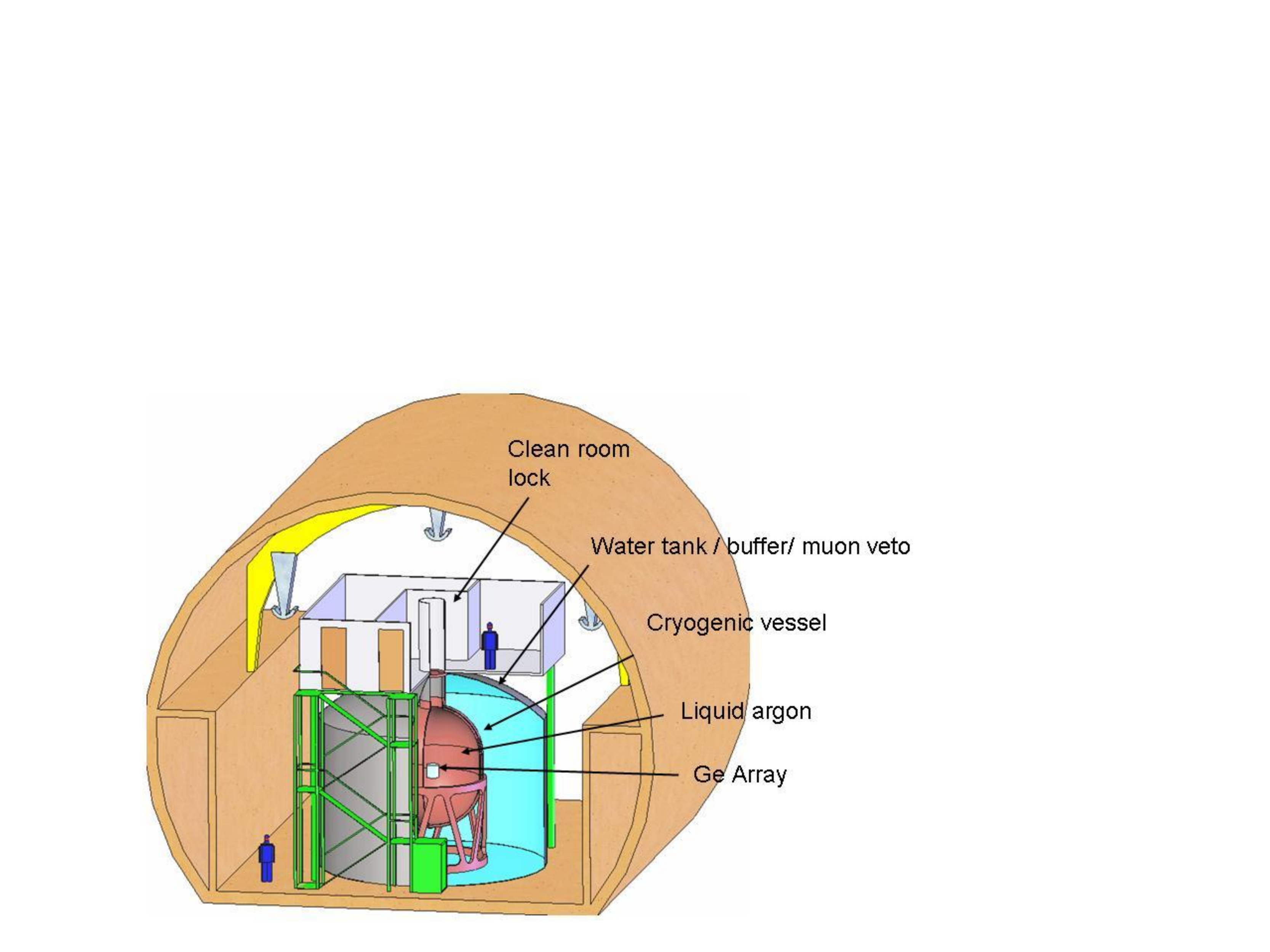}
\caption{\label{fig:baseline} An artist's view of the \gerda baseline design.}
\end{center}
\end{figure}

The conceptual design of the \gerda experiment is based on an earlier proposal \cite{heu} to submerge naked Ge detectors inside a large volume cryogenic fluid shield. This concept is based on the observation that the background signals are largely dominated by external radiation.  Naked Ge detectors will be immersed directly in the liquid argon. The cryogenic volume is surrounded by a buffer of ultra-pure water acting as an additional gamma and neutron shield as depicted in an artist's view of the \gerda baseline design displayed in Fig. \ref{fig:baseline}.

\begin{figure}[t]
\centering
\subfigure[The expected 90~\% probability lower limit on the half-life for neutrinoless double beta decay is plotted versus the exposure under different background conditions.] 
{
    \label{fig:2a}
    \includegraphics[width=5.5cm]{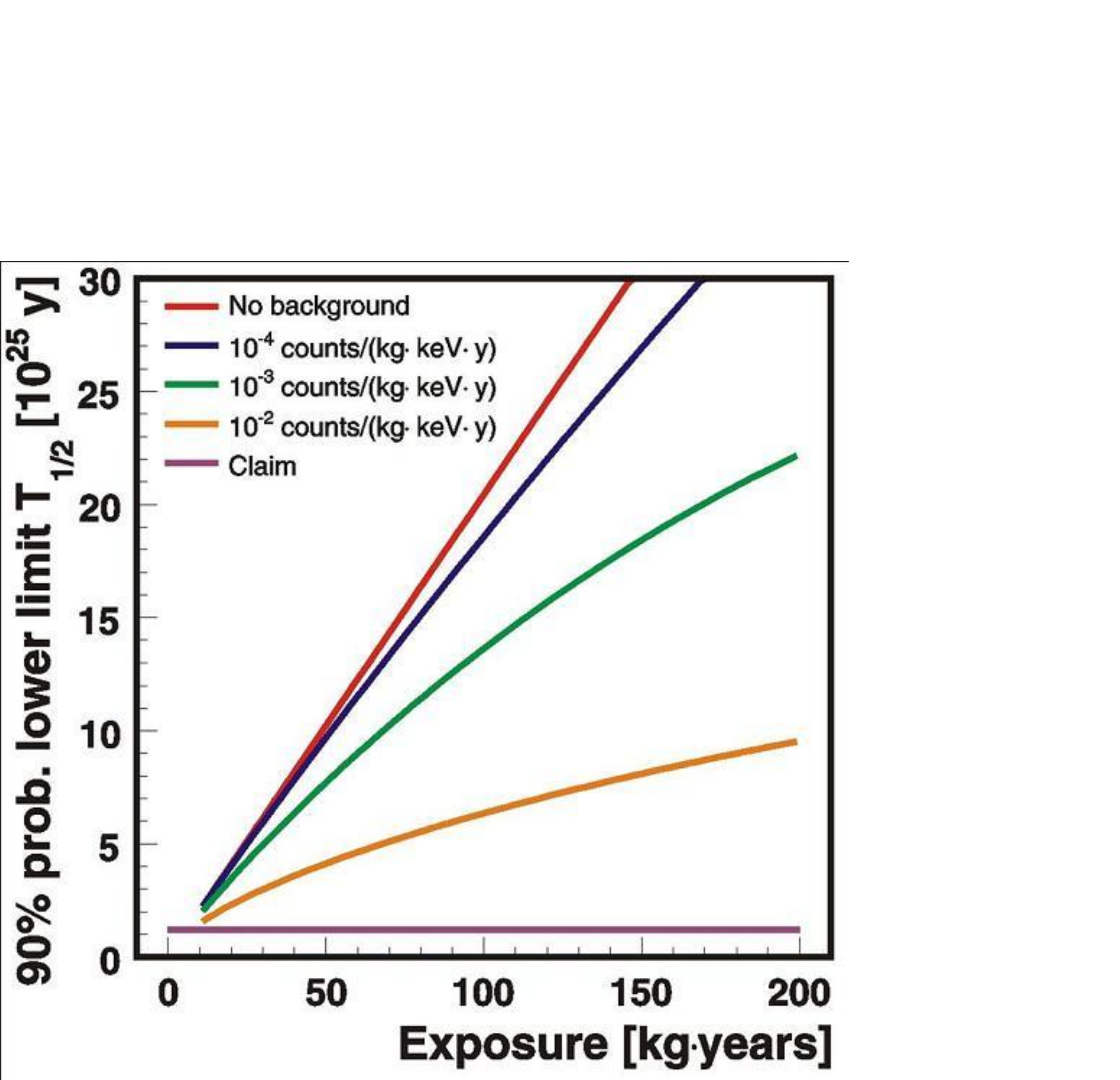}
}
\hspace{1cm}
\subfigure[The upper bounds that can be put on the effective neutrino mass with $90\%$ probability assuming matrix element value $<M^{0\nu}>=3.92$ \cite{correction}.] 
{
    \label{fig:2b}
    \includegraphics[width=6.0cm]{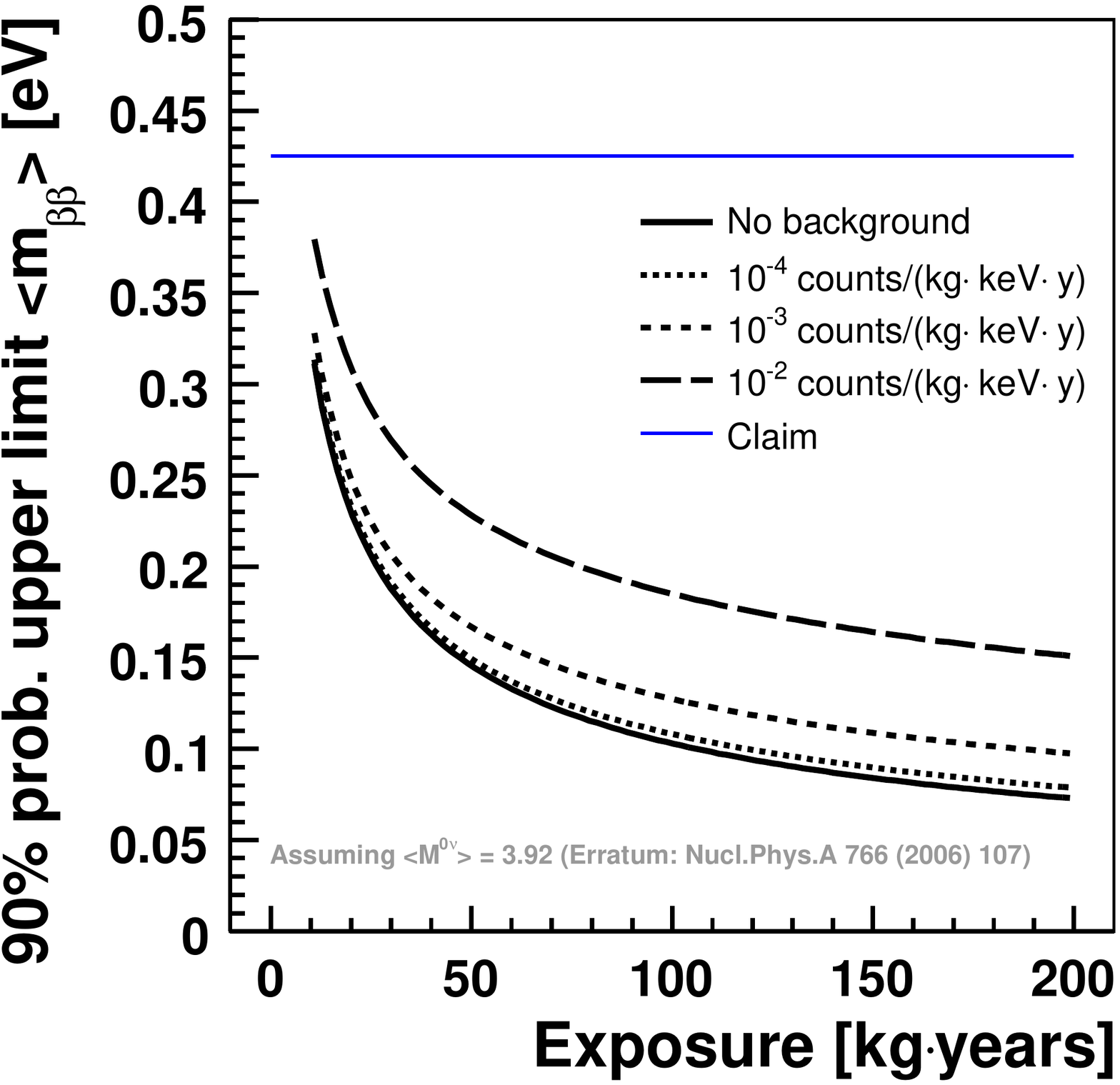}
}
\caption{\label{fig:sub} Expected sensitivities depending on exposure and background conditions. The claimed observation for neutrinoless double beta decay of $^{76}$Ge \cite{claim} is also shown.} 
\end{figure}

The sensitivity obtainable in double beta decay experiments with a given exposure and background index was calculated using Monte-Carlo ensemble test on the basis of Bayesian statistics \cite{PhysRevD74092003} and is displayed in Fig. \ref{fig:2a} and Fig. \ref{fig:2b}. The $90\%$ prabability for obtaining lower limits on $T_{1/2}$ higher than the displayed values as a function of exposure for given background index are given in Fig. \ref{fig:2a}. Fig. \ref{fig:2b} shows the upper bounds that can be put on the effective neutrino mass with $90\%$ probability using matrix element value from \cite{correction}.

The phased approach of \gerda and the physics reach are the following: The \gerda experiment will be carried out in several phases with each of them providing relevant physics results. In Phase I of \gerda the existing 18 kg of enriched germanium detectors, which have been used in the IGEX and Heidelberg-Moscow experiment, will be refurbished and operated as bare detectors in liquid argon. After two years of data taking, corresponding to an exposure of 30 kg$\cdot$years, we can either confirm the claimed observation \cite{claim} of neutrinoless double beta decay or refute it as can be seen in Fig. \ref{fig:2a} and Fig. \ref{fig:2b}. If the claim is correct, for ($0\nu\beta\beta$) decay of $^{76}$Ge with T$_{1/2} = 1.2\cdot 10^{25}$y about 13 events above a background of 3 events within 10 keV window around Q$_{\beta\beta}$ are expected after 2 years of measurement. If no events will observed, the limit on the half life would amount to T$_{1/2} > 3\cdot 10^{25}$y (90\% C.L.) or, tranlated into an effective neutrino mass, m$_{\beta\beta}<0.3 - 0.9$eV, depending on nuclear matrix element used (for instance, using the corrected matrix element $<M^{0\nu}>=3.92$ from \cite{correction} an upper limit of 0.27 eV can be set with 90$\%$ probability).

For Phase II about 22 kg of new $^{76}$Ge detectors will be added and the total mass will be approximately 40 kg. The new detectors will be of true coaxial geometry with segmented electrode readout enhancing the performance of discrimination of single-site from multi-site events. After data taking corresponding to an exposure of 150 kg$\cdot$years, and assuming a background will reduced up to $10^{-3}$counts/(keV$\cdot$kg$\cdot$y), the limit on T$_{1/2}$ would improve to $>1.5\cdot 10^{26}$y (90$\%$ C.L.). This translates to an upper limit on the effective neutrino mass of 0.09 - 0.29 eV (0.11 eV assuming $<M^{0\nu}>=3.92$ from \cite{correction}).

\begin{figure}[h]
\begin{center}
\includegraphics[width=0.75\linewidth]{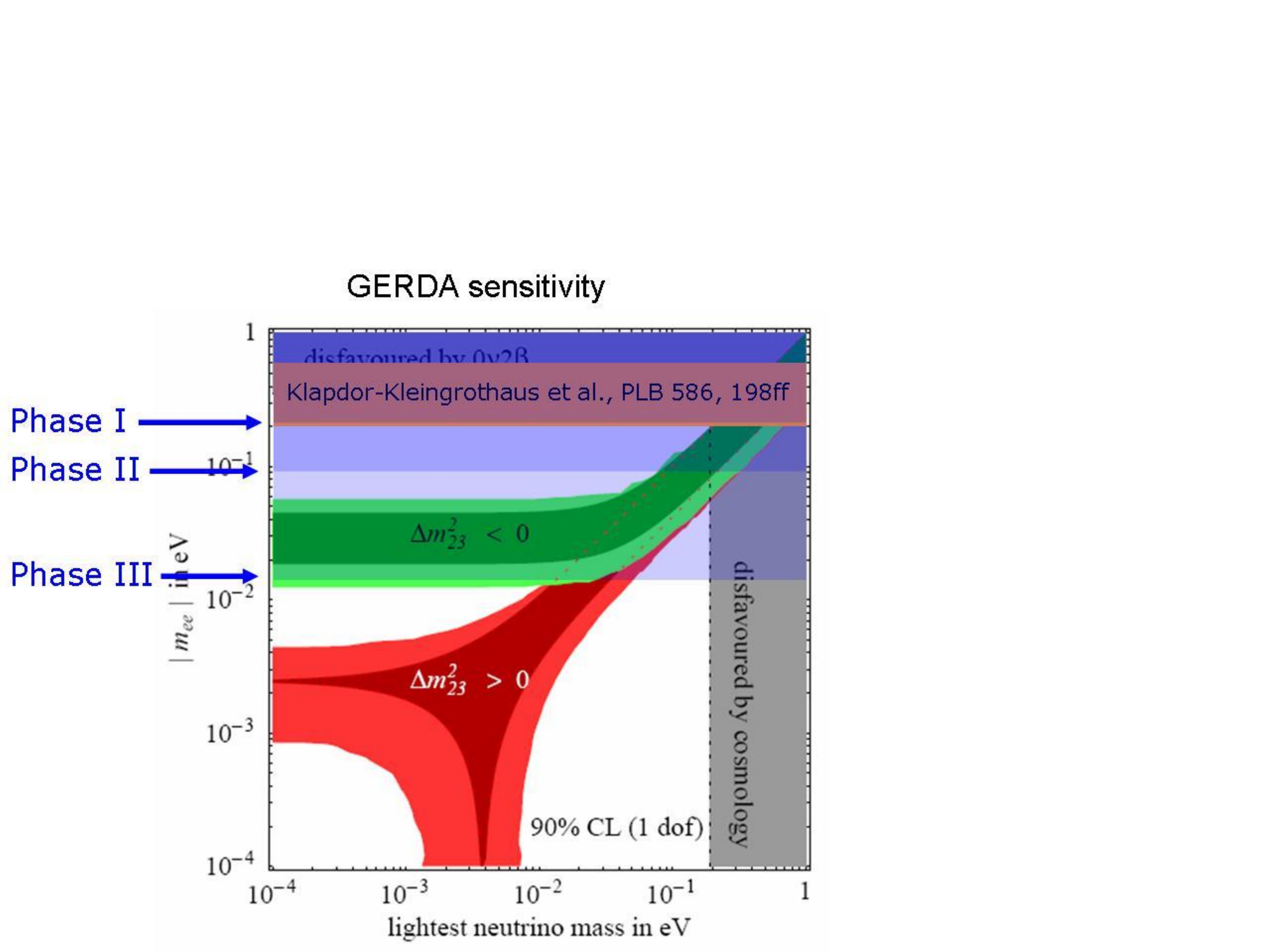}
\caption{\label{fig:3} A plot depicting the sensitivity of the three phases of GERDA for the scale of the effective Majorana electron neutrino mass. The region printed in green is the inverted hierarchy, the normal hierarchy region is shown in red.}
\end{center}
\end{figure}

Figure \ref{fig:3} shows a plot of the allowed regions of the effective Majorana electron neutrino mass against the lightest neutrino mass \cite{FSV03}. In the plot the areas corresponding to the sensitivity of the two phases of \gerda are visualized along with the claim of detection mentioned above. Phase I will cover the area of sensitivity required to scrutinize the claim and Phase II will cover the degenerate neutrino mass hierarchy. If no signal for the (0$\nu\beta\beta$) decay is found, a ton scale $^{76}$Ge experiment (Phase III) with further background reduction up to $10^{-4}$counts/(keV$\cdot$kg$\cdot$y) undertaken in a worldwide collaboration will be required to cover the inverted hierarchy region.

\section{GERDA setup}
\begin{figure} [h]
\begin{center}
\includegraphics[width=0.6\linewidth]{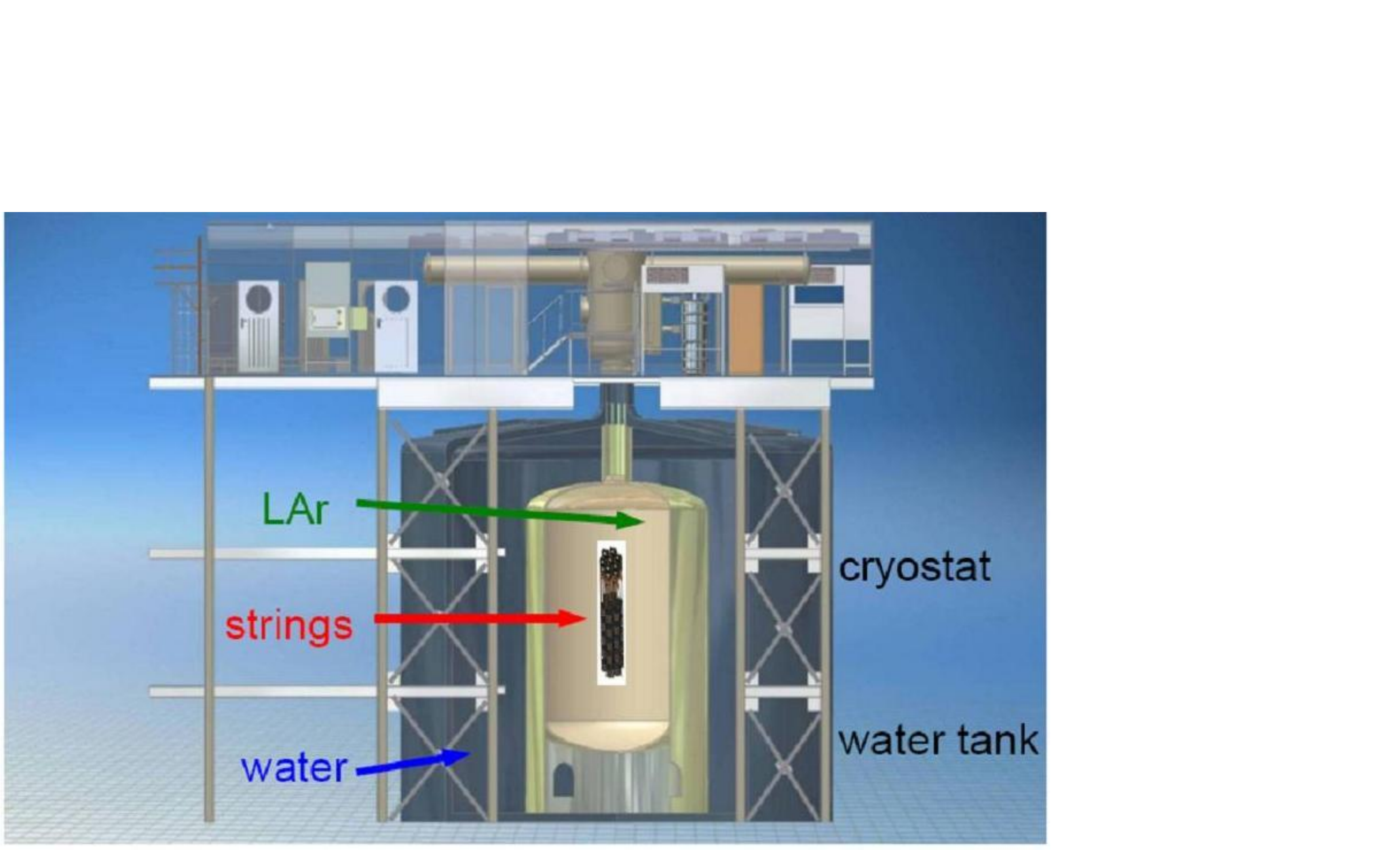}
\caption{\label{fig:4} A schematic drawing of the complex GERDA set up.}
\end{center}
\end{figure}

A schematic drawing of the complex \gerda setup is shown in Fig. \ref{fig:4}.

\begin{figure} [h]
\begin{center}
\includegraphics[width=0.6\linewidth]{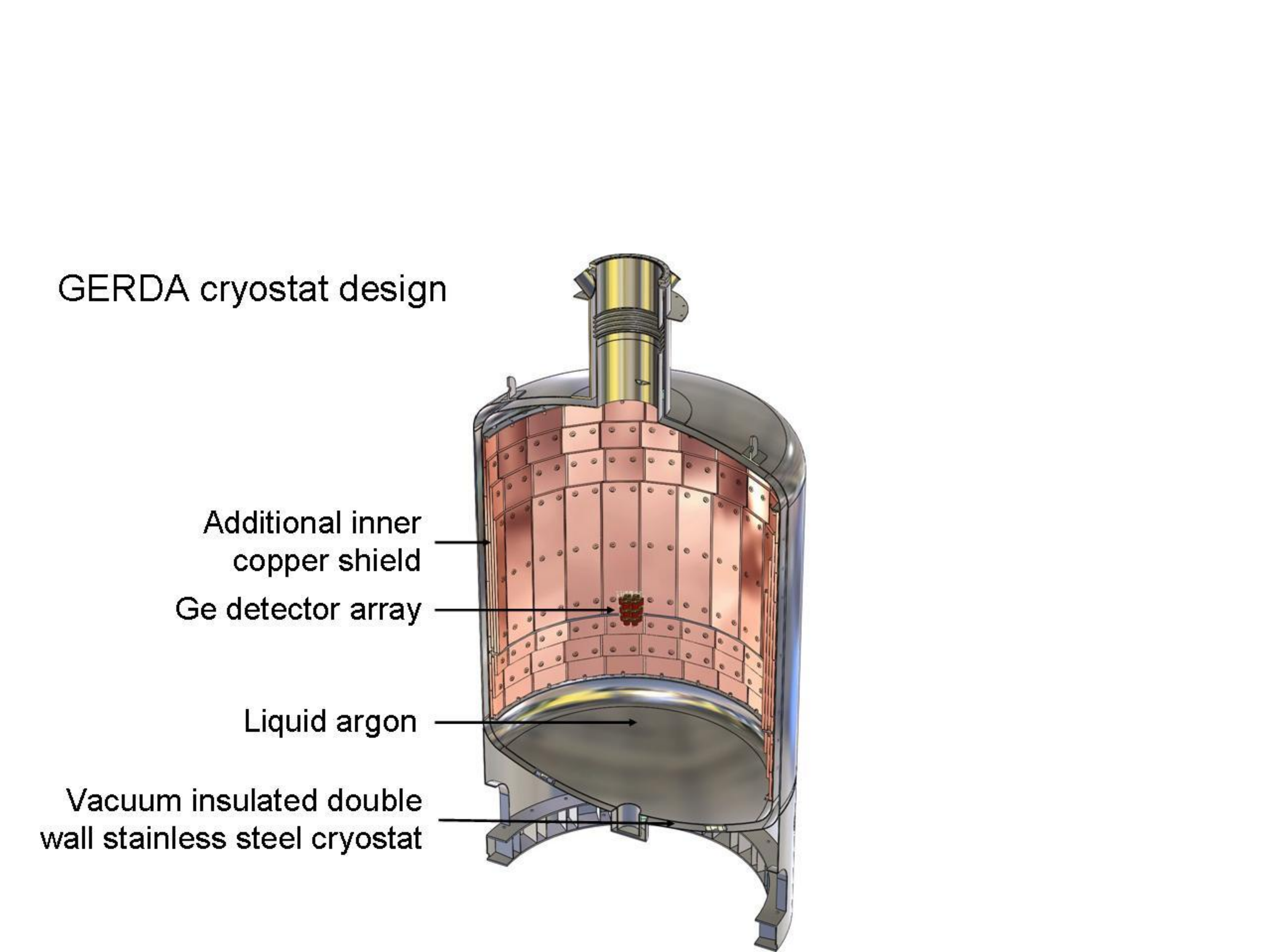}
\caption{\label{fig:5} The superinsulated stainless steel cryostat with internal high-purity copper shield.}
\end{center}
\end{figure}

\begin{figure} [h]
\begin{center}
\includegraphics[width=0.4\linewidth]{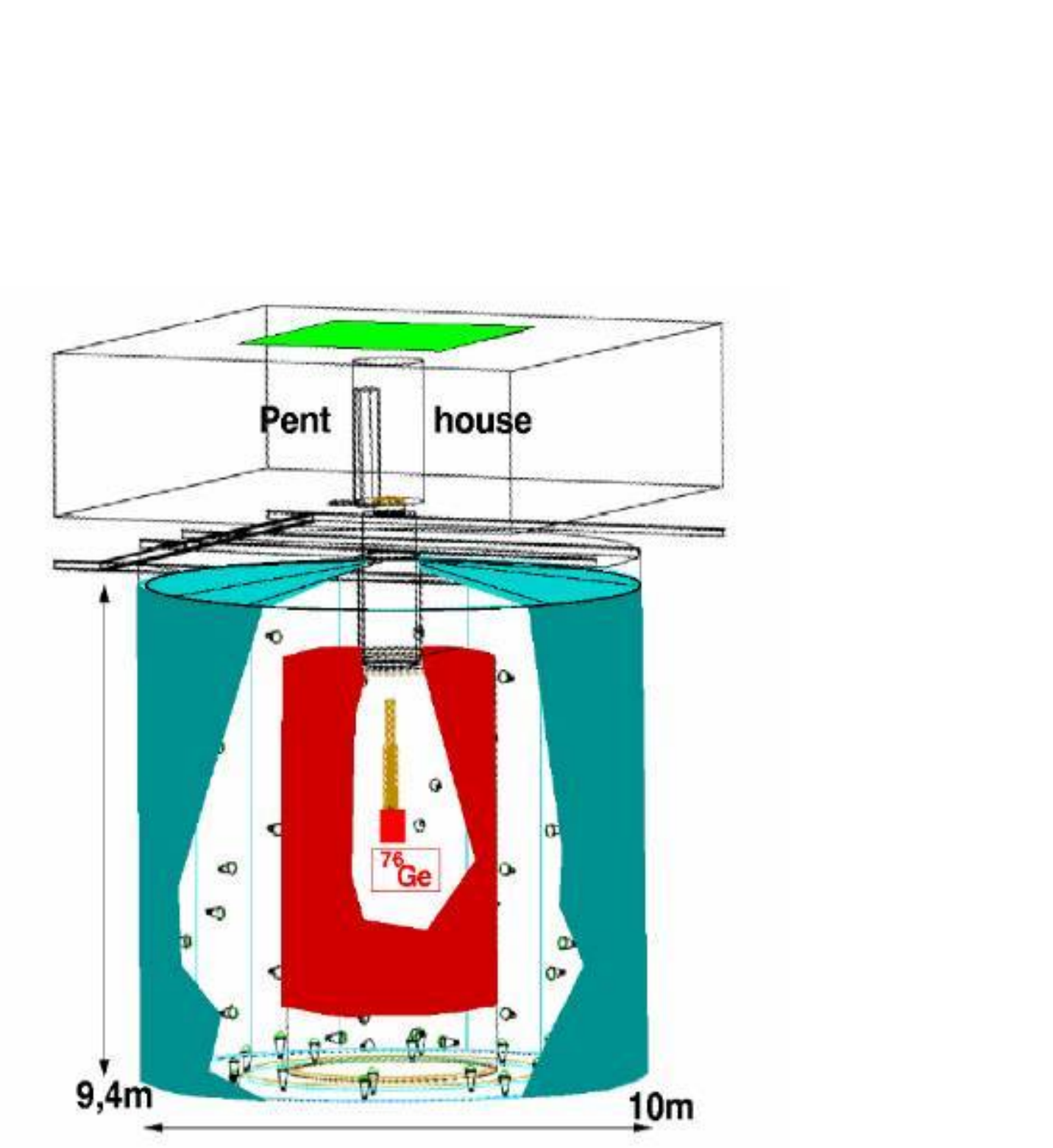}
\caption{\label{fig:6} Water Cherenkov detector instrumented with 66 photomultipliers.}
\end{center}
\end{figure}

A vacuum insulated stainless steel cryostat of 4.2 m diameter and total height (including the neck) of 8.9 m with the inner vessel volume 70 m$^3$ will contain 98 tonnes of LAr. An inner cylindrical shell of the cryostat is covered by internal ultrapure OFE Cu shield with maximum thickness of 6 cm and total mass of about 20 tonnes (see Fig. \ref{fig:5}). The cryostat is immersed in a water tank with a diameter of 10 m and a total height of 9.4 m. The water buffer serves also as a gamma and neutron shield and, instrumented with 66 photomultipliers, as Cherenkov detector for efficiently vetoing cosmic muons  (Fig. \ref{fig:6}). Plastic scintillator panels on top of the detector with the total square of about 20 m$^2$ will tag muons which enter the dewar through the neck.

\begin{figure}[h]
\centering
\subfigure[5x7 Ge detector array assembled from 7 strings of 5 detectors] 
{
    \label{fig:7a}
    \includegraphics[width=3cm]{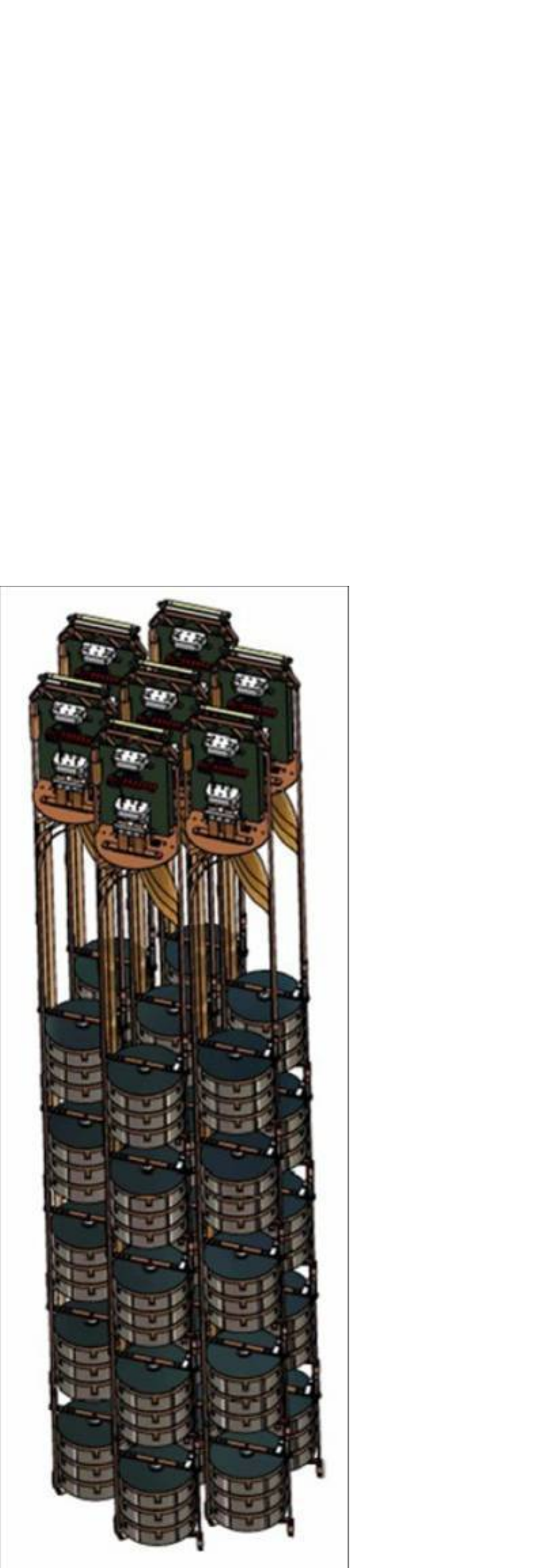}
}
\hspace{0.5cm}
\subfigure[p-type Ge detector for Phase I] 
{
    \label{fig:7b}
    \includegraphics[width=4cm]{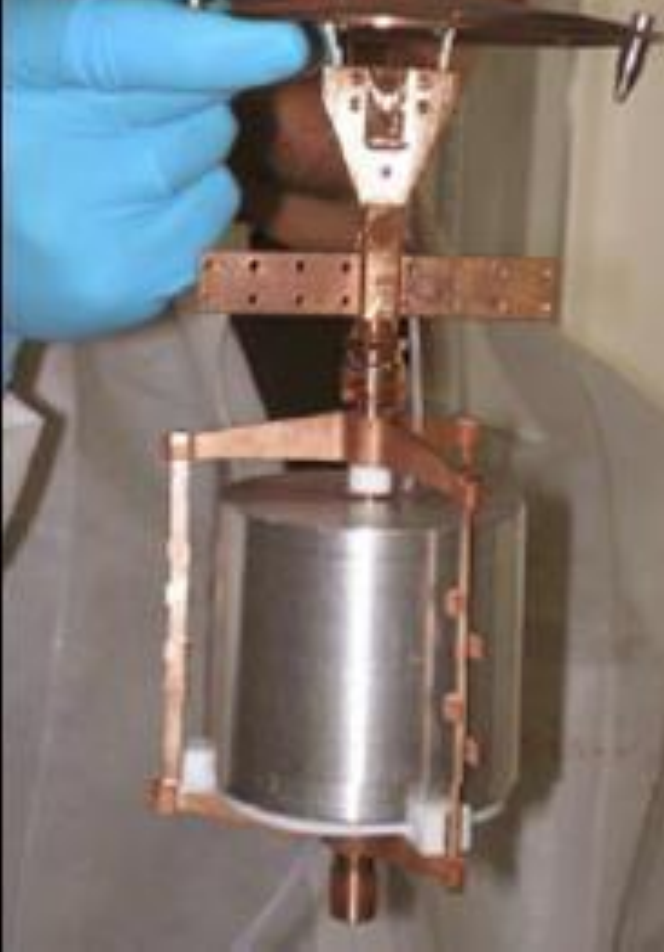}
}
\hspace{0.5cm}
\subfigure[segmented true coaxial n-type Ge detector for Phase II in low mass support and contact structures.] 
{
    \label{fig:7c}
    \includegraphics[width=4cm]{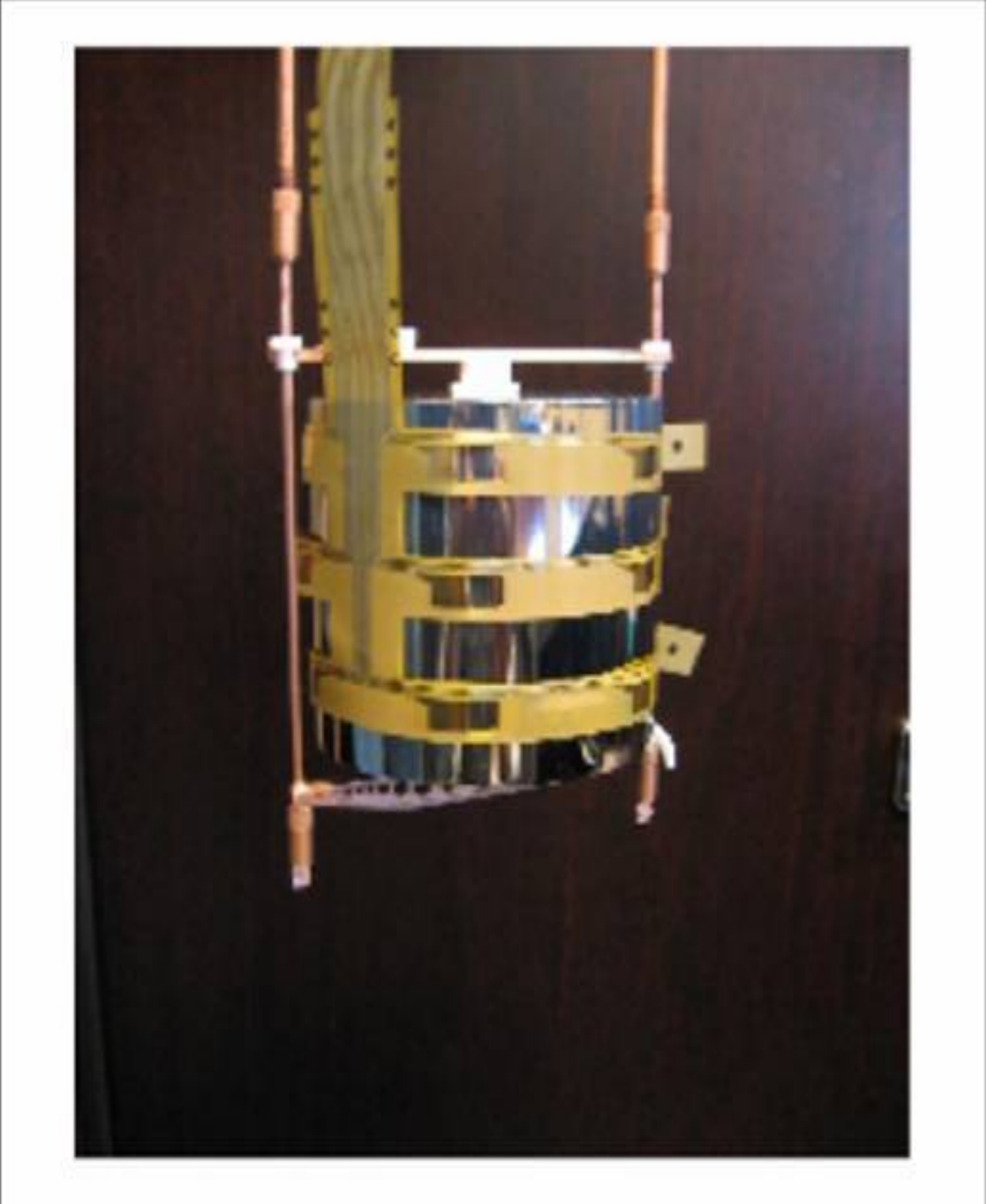}
}
\caption{\label{fig:7all} Ge detector array and single detectors.} 
\end{figure}

The Ge detector array, shown in Fig. \ref{fig:7a}, has a hexagonal structure and is made up of individual detector strings. A cleanroom and radon tight lock on top of the vessel assembly allow to insert and remove individual detector strings without contaminating the cryogenic volume.  Calibration sources can be inserted close to the detectors. A detector string is assembled from up to five independent Ge detector modules. Designs of such modules are shown in Fig. \ref{fig:7b} and \ref{fig:7c} for p-type (Phase I) and segmented n-type Ge diodes (Phase II).

\begin{figure}[h]
\begin{center}
\includegraphics[width=0.5\linewidth]{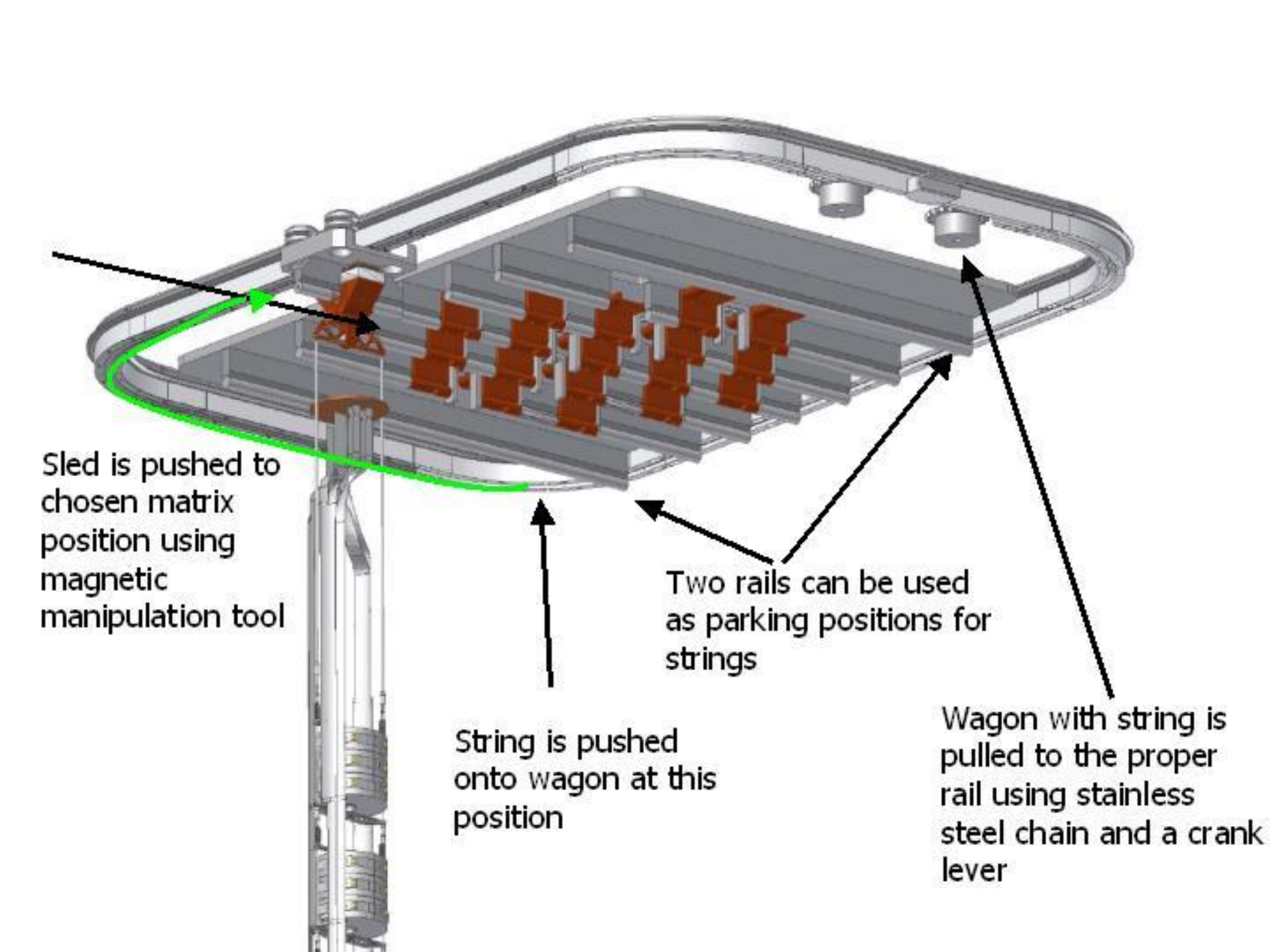}
\caption{\label{fig:8} The rail system to position the detector strings in the array.}
\end{center}
\end{figure}

The upper infrastructural complex is sitting on the superstructure. Its main constituent is the clean-room which houses the lock-system for insertion of the detectors to the cryogenic volume. The main parts of the lock system are the inner and the outer lock as well as two cable tubes. The inner lock houses the rail system to position the detector strings in the array (Fig. \ref{fig:8}). 

The cable tubes contain the linear pulleys to lower individual detector strings to the cryogenic volume. The inner lock system can be decoupled from the cryogenic tank with a circular shutter. Rectangular shutters separate the inner from the outer lock and the outer lock from the clean-room. The lock system is supported by steel bars that rest on the superstructure.

\clearpage
\section{The Phase I and Phase II detectors}

All IGEX and HdM detectors have been characterized and tested, the same energy resolutions as previously were obtained. After they were removed from their cryostats, are presently being refurbished and will then be installed into the Phase I suspension. A clean underground detector test facility has been installed at LNGS for detector R\&D. A successful refurbishment procedure has been developed and verified for the HdM and IGEX detectors. Moreover, optimum procedures are being developed for the operation of bare Ge detectors in LAr with one of the prototype detectors (from non-enriched Ge); both the long term stability is investigated as well as the cycling between room and LAr (88K) temperature.  The prototype detector has been operated in liquid argon in the holder at LNGS since beginning of 2006, it has gone through more than 40 cooling and warming up cycles (see \cite{barnabe}). The enriched materials for phase II of the experiment, 35.5 kg of germanium enriched to 87\%-88\% in $^{76}$Ge in form of GeO$_2$, have been produced in Russia and were transported from Siberia to Germany in a steel cylinder designed to reduce cosmogenic activation. The material is now stored in a 500 m.w.e. underground site. The phase II detectors will be segmented true coaxial n-type, 18-fold in the presently working prototype design. The segmentation will help to identify multiple Compton scattering events as background ones in the region of interest with high efficiency depending on its source and location.
Figure \ref{fig:7c} shows an 18-segment prototype detector (3-fold along $z$ and 6-fold along $\phi$) mounted in its copper suspension and with a prototype kapton cable attached. The suspension consists of only 30 g copper and 7 g Teflon. The purity requirements on these materials are stringent, as they are very close to the detectors. The 18 segments are read out using a novel contacting scheme described in \cite{Abt07}. The prototype detector was extensively tested in a conventional test cryostat. The energy resolution of all segments and the core were around 3 keV at 1.3 MeV.

\section{Background sources and suppression methods}

Two general classes of the background usually play role in such type of the experiments: external and internal backgrounds. The principal concept of the \gerda experiment foresees a possibility to suppress considerably both of them by using a powerful multilevel passive shield with the ultra pure liquid argon as its inner part, and by using as minimal as possible cladding materials around the bare Ge crystals.

The 3~m thick layer of highly purified water plus the 2~m thick cryogenic shield must reduce the radioactivity of the surrounding rock and concrete (3 Bq/kg $^{228}$Th)  to the desired background index of a few 10$^{-4}$ counts/(keV$\cdot$kg$\cdot$y).

To prevent the additional contribution from the construction materials between the water shield and Ge detectors, only selected materials with extremely low radioactive contamination are used.  The special stainless steel used for the cryostat production (3~cm double wall) has been found with contamination of $^{228}$Th and $^{226}$Ra $<$ 1 mBq/kg, that gives negligible contribution to the total external background: $<1.7\cdot 10^{-5}$ counts/(keV$\cdot$kg$\cdot$y) and $<2.3\cdot 10^{-6}$ counts/(keV$\cdot$kg$\cdot$y), respectively. Radioactive contamination of the inner Cu shield (3 - 6 cm thickness) is negligible even in comparison with the cryostat steel ($^{228}$Th $<39~\mu$Bq/kg; $^{226}$Ra $<50~\mu$Bq/kg). Superinsulation will contribute even less than stainless steel, $^{228}$Th $<2.2\cdot 10^{-6}$ and $^{226}$Ra $<1.4\cdot10^{-6}$ counts/(keV$\cdot$kg$\cdot$y).
Several samples of Liquid Ar produced by different companies have been tested for radon contamination. It was found that initial $^{222}$Rn activity is in the range from 0.4 to 4 mBq/m$^3$ (STP), depending of the companies and methods of producing. Required limit on $^{222}$Rn concentration corresponds activity less than 0.5 µBq/m$^3$ with contribution to background $<10^{-4}$ counts/(keV$\cdot$kg$\cdot$y). It means that purification of LAr from $^{222}$Rn with a factor of about 1000 is needed.

Careful measurements and checking of radiopurity of all materials to be used for \gerda are extensively performed with using of the low-background Ge-spectrometers, (see \cite{Budjas}), inductively coupled plasma mass spectrometry, neutron activation, Rn emanation methods, surface  alpha activity counting and other relevant methods. More than hundred samples have been assayed by the collaboration during the last two years.  Most recent  gamma-spectrometry screening results have been obtained for all ($> 10$) relevant batches of the 30 tons of 1.4571 steel from which the cryostat will be fabricated. As a result, material with less than 1 mBq($^{228}$Th)/kg activity has been identified for the cryostat's cylindrical shells.

Active shield consisted from the powerful water Cherenkov detector and the plastic scintillator array must reduce considerably the other part of the external background produced by cosmic rays up to the level $<10^{-4}$ counts/(keV$\cdot$kg$\cdot$y).

The internal background of the Ge detectors is mostly due to cosmogenic spallation in germanium, radioactive contamination in surrounding materials (detector support, cabling and electronics), surface contamination of the Ge crystals, and neutron activation. The most contribution comes from cosmogenic $^{68}$Ge and $^{60}$Co, since their lifetimes are in the order of years and their decays can produce events in the region of interest. In this connection it is required to reduce the exposure of the enriched Ge material to cosmic rays as much as possible.

The active methods of the internal background reduction developed for the Phase I and / or Phase II are the following: 1) anti-coincidence between detectors; 2) segmentation of the Ge detectors (Phase II); 3) pulse shape analysis; 4) coincidence in decay chain ($^{68}$Ge); 5) scintillation light detection (LArGe test facility, Phase II). More probably, all these techniques will be necessary to apply in order to reach the specified background levels less than 10$^{-3}$counts/(keV$\cdot$kg$\cdot$y) for Phase II.

First three from the above mentioned methods are exploit that beta decay has a point-like energy deposition inside Ge detector because of  the range of electrons in germanium is of the order 1 mm and the energy deposition of the (0$\nu\beta\beta$)-signal is typically strongly localized inside the diode. The $^{60}$Co and $^{68}$Ge decays lead mostly to extended multi-site events due to Compton interactions of gamma rays with energies in the 2 MeV range. Such multi-site events will be suppressed by anti-coincidence of detectors within the array or, due to higher granularity even more efficiently, in segmented detectors. For $^{60}$Co and $^{68}$Ge inside segmented detectors, the suppression factor is more than an order of magnitude.

\begin{figure}[h]
\begin{center}
\includegraphics[width=0.7\linewidth]{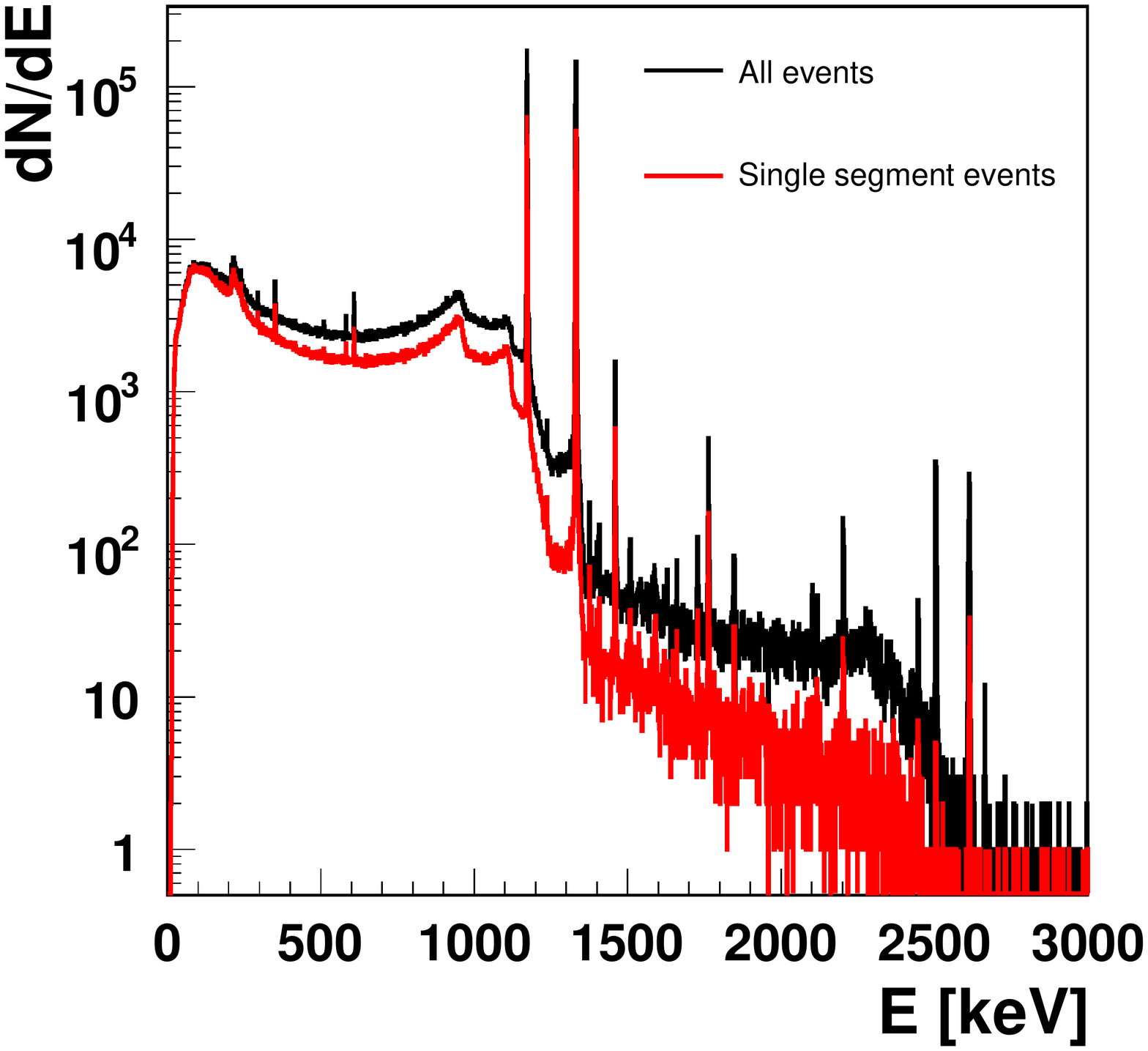}
\caption{\label{fig:9} Energy spectrum accumulated from the core electrode taken with a $^{60}$Co source 5 cm above the crystal (black line). The resultant spectrum when it is required that only one segment has an energy deposition above 20 keV is also shown (red line). At around 2MeV about $90\%$ of the events are rejected \cite{segmentationDATA}.}
\end{center}
\end{figure}

Recently obtained data were taken with one of the \gerda Phase II prototype detector. It has a mass of approximately 1.6~kg and an 18-fold segmentation (6-fold in the azimuthal angle $\phi$ and 4-fold in the height $z$). The ability to identify photon induced events was studied in detail with several sources, after background subtraction the suppression of events in the $Q_{\beta\beta}$-region was measured to vary between 2 and 100, depending on the source~\cite{segmentationDATA}. To demonstrate the feasibility of the identification in Fig. \ref{fig:9} a core spectrum taken with a $^{60}$Co source 5 cm above the crystal is shown. Also shown in Fig. \ref{fig:9} is the resultant spectrum when it is required that only one segment has an energy deposition above 20keV. At around 2MeV about $90\%$ of the events are rejected. A detailed analysis is presented in \cite{segmentationDATA}.

Another effective approach is to discriminate multi-site deposits from the pulse shape analysis of the signal. This is possible due to a correlation between the location of the energy deposition and the relative drift-times for the electrons and holes in the radial electric field applied to the Ge diode . As a result the pulse shape of signals from multiple energy deposition sites is in the most case different from the signal of a single site event. This method has been previously used in the HdM and IGEX experiments.

Coincidence in decay chain can be used to select background events which are intrinsically correlated to subsequent decays with a specific energy signature. For example, in the $^{68}$Ge decay an electron capture is followed with a half life of T$_{1/2}$ = 68 min by the $\beta^+$ decay of $^{68}$Ga, which can produce a background candidate. The time correlation can be used to veto these events.

Background discrimination by the liquid argon scintillation readout of events coincident with Ge detector signals would be orthogonal with respect to pulse shape and segmentation methods. More detailed description of this method is given below in chapter 6.

The $^{222}$Rn concentration in the air of the Gran Sasso laboratory is variated between 50 and 120 Bq/m$^3$. For the handling of germanium crystals this concentration is prohibitively high. Therefore we consider the possibility to supply radon-reduced air to the clean room on top of the tank as well as radon measurement devices to monitor the activity. Radon can be removed from air by adsorption on an activated carbon column.

Monte Carlo studies \cite{Pan07} were performed to estimate the background expected for Phase II of the \gerda experiment. It was shown, that with the ongoing work on further material selection and  a progress in development of new methods of background suppression the goal of a total background index $\leq 10^{-3}$ counts/(keV$\cdot$kg$\cdot$y) for Phase II can be reached.

\clearpage
\section{LAr scintillation study and the LArGe test facilities}

For Phase I the required background can be reached with passive shielding and existing background suppression techniques mentioned above. To reach the background level required for Phase II, new background suppression methods have to be implemented.  A novel method was studied to suppress backgrounds in the later \gerda phases. LAr scintillates in UV ($\lambda$=128~nm) and the main concept is the simultaneous readout of the signal from the Ge detector (immersed in LAr)  and the scintillation light of the LAr, which is then used as anti-coincidence signal.  Compton scattered gammas inside the germanium crystals can deposit part of their residual energy in the liquid argon volume and this class of events can be vetoed.

\begin{figure}[h]
\begin{center}
\includegraphics[width=0.5\linewidth]{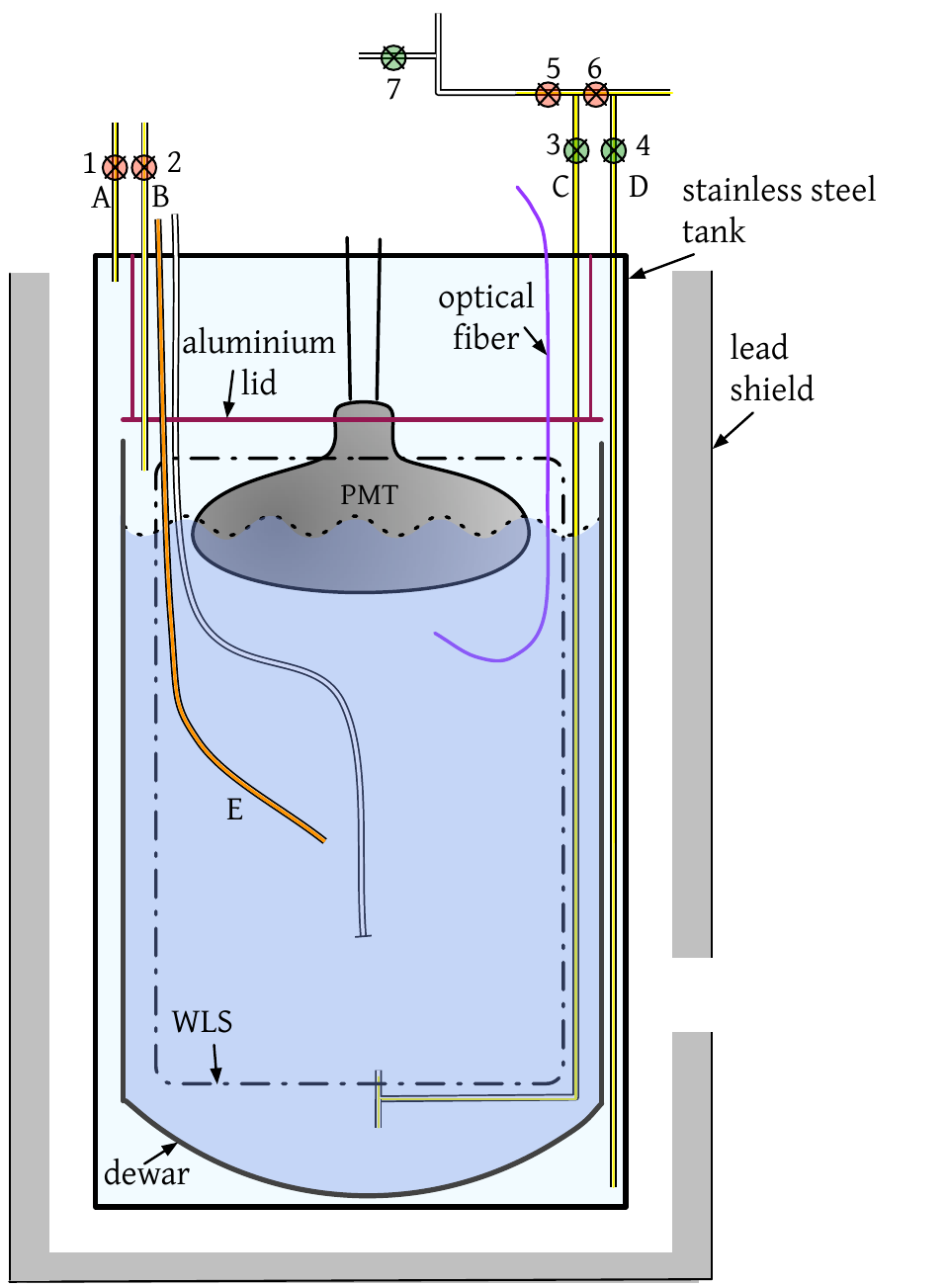}
\caption{\label{fig:10} A schematic drawing of the medium size mini-LArGe set up.}
\end{center}
\end{figure}

In the framework of the GERDA R\&D program the efficiency of such a LAr scintillation veto was investigated for the first time. For this purpose the first Liquid Argon-Germanium hybrid detector has been developed and successfully operated in the low level underground laboratory at the MPI-K in Heidelberg. For this medium size test set up ("mini-LArGe") the active LAr volume was chosen as 13.5 l (h=43 cm, diameter =20 cm) which corresponds to 19 kg of LAr (Fig. \ref{fig:10}). In this test system a 390 g p-type HP-Ge diode was operated, suspended in the LAr.

Specially designed cryogenic photomultipliers (ETL 9357-KFLB) are used to detect LAr scintillation light. Since the glass of phototubes is not transparent to the extreme ultra violet (XUV) scintillation-light it has to be shifted before detection. For this purpose a photocathode window is coated by a wavelength-shifter (WLS) and the active volume is surrounded by a foil which acts both as WLS and as reflector in the optical range. The wavelength-shifter absorbs the XUV light and emits in the optical range ($\lambda$=410-450 nm).

\begin{figure}[h]
\begin{center}
\includegraphics[width=0.75\linewidth]{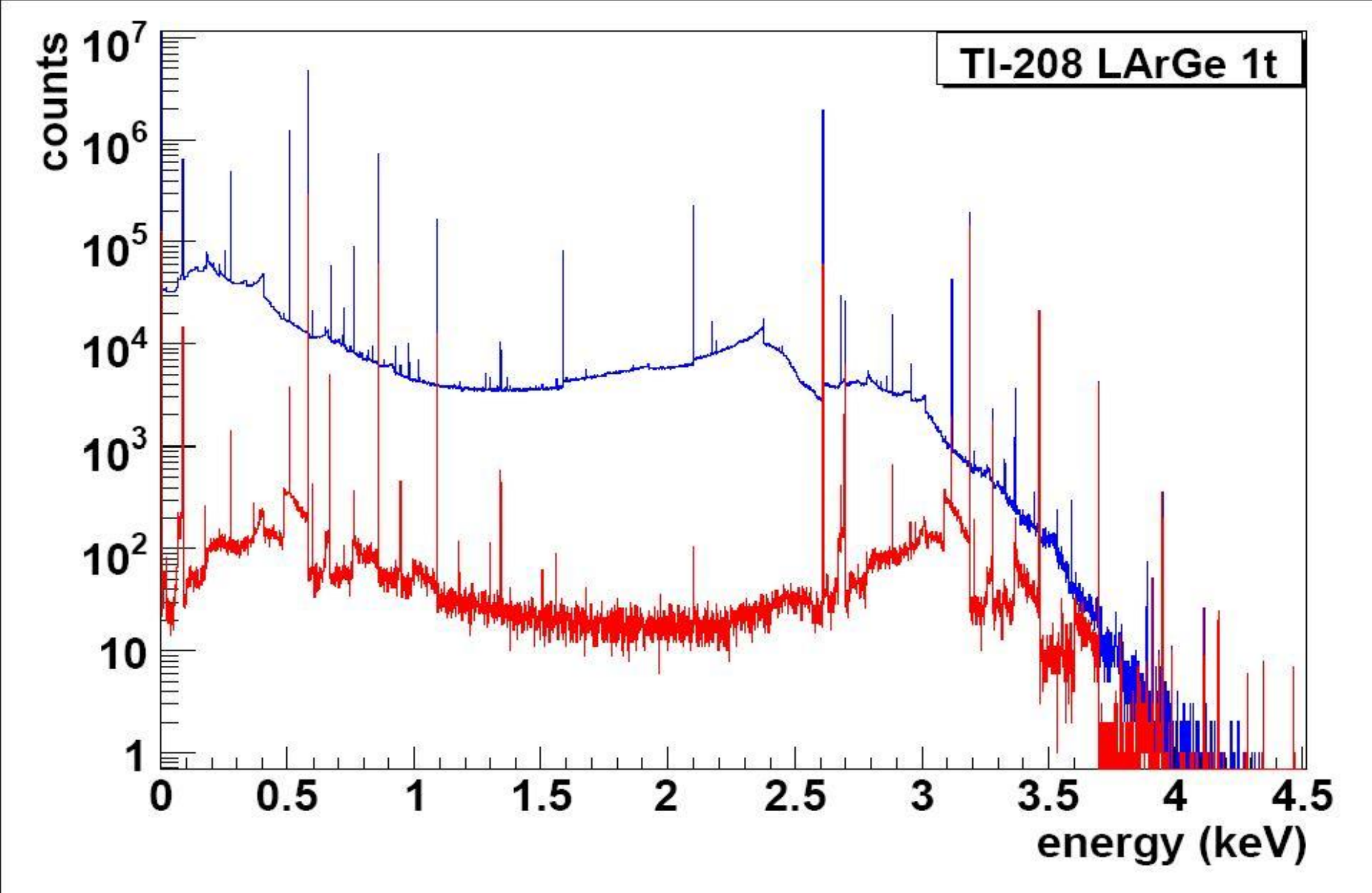}
\caption{\label{fig:11} The Monte Carlo simulation of the suppression efficiency in the LArGe setup at the Gran Sasso underground laboratory (1 ton of active LAr volume) for $^{208}$Tl source. The unsuppressed spectra in the Ge diode are plotted in blue, the suppressed spectra are plotted in red. The suppression factor predicted for the RoI is equal to 338$\pm$8.}
\end{center}
\end{figure}

Even for this small size set up, the power of the LAr scintillation anticoincidence concept for background suppression has been successful demonstrated. With an active volume of 19 kg a suppression factor of 3 for $^{60}$Co and of 17 for $^{232}$Th has been reached in the region of
interest around Q$_{\beta\beta}$=2039 keV. The corresponding Monte Carlo simulation are in reasonable agreement with the experimental data. In this set up the suppression factors are limited by escaping $\gamma$-s from the limited active LAr volume (10 cm radius). For a new LArGe setup with  larger active volume (h = 200 cm, diameter =90 cm) the escape probability becomes negligible and the performed Monte Carlo simulations predict suppression factors of about 100 for $^{60}$Co and of 340 for $^{208}$Tl. (see Fig. \ref{fig:11})

The LAr scintillation was also investigated as a tool for diagnostics and selection of different background components (gamma, neutron, alpha) by using the opportunity of pulse shape analysis. To achieve this goal with considerable discrimination factor the light yield even higher than for pure veto purposes is required. For this purpose a new, very stable and robust wavelength shifter/reflector has been developed, leading to a photo electron (pe) yield of as much as 1200 pe/MeV.

\begin{figure}[h]
\centering
\subfigure[normalized entire pulse shapes (PDF)] 
{
    \label{fig:12a}
    \includegraphics[width=0.45\linewidth]{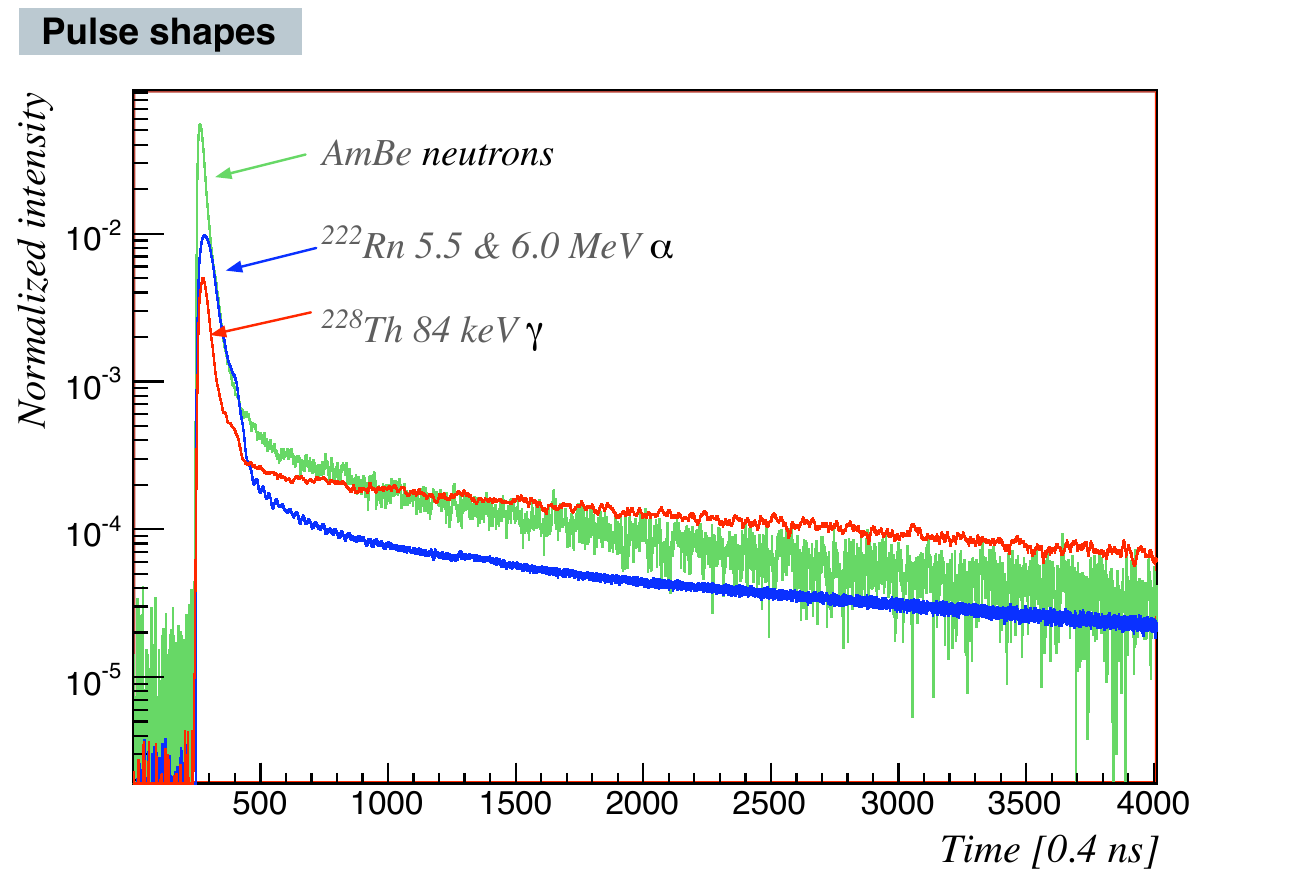}
}
\hspace{0.5cm}
\subfigure[cumulative distribution functions (CDF)] 
{
    \label{fig:12b}
    \includegraphics[width=0.45\linewidth]{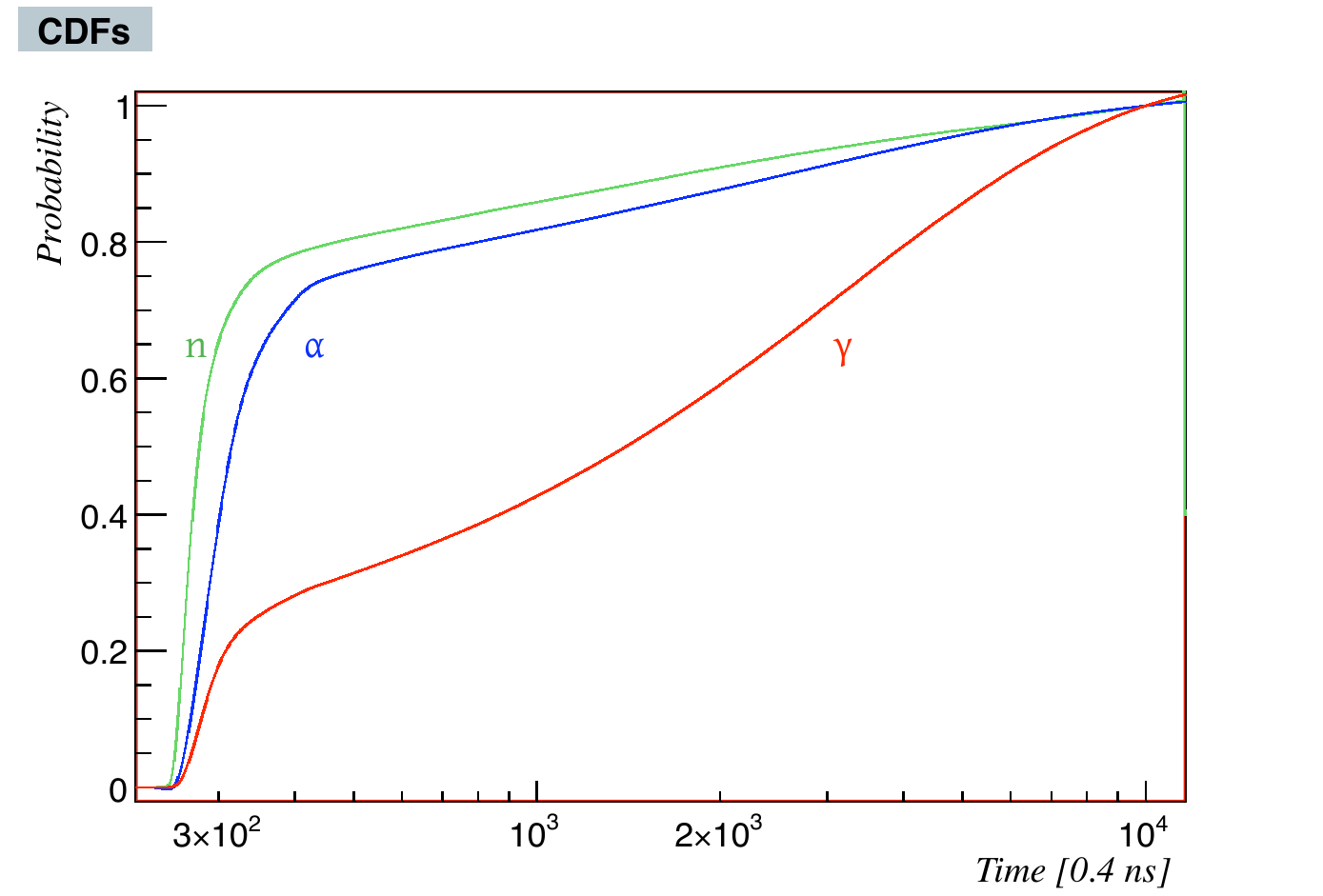}
}
\caption{\label{fig:12all} Average pulse shapes of gamma, alpha and neutron radiation.} 
\end{figure}

The possibility to perform pulse shape discrimination with using LAr scintillator is based on the following concept. The scintillation light of argon is emitted from excited dimers (excimers) which are created in argon by ionizing radiation. These excimers can be created in a singlet or a triplet state. The de-excitation times of these states are $\tau_s$=6 ns and $\tau_t=1.6 \mu$s respectively and their population depends on the ionization density of particle \cite{KSR80}. This allows discrimination between particles with different ionization densities by analyzing the pulse shape of the LAr scintillation. This means that the time dependence of the scintillation light intensity I(t) is dominated by two exponential decays. One with the time constant of the singlet state de-excitation $\tau_s$ (6 ns), the other with that of the triplet state de-excitation $\tau_t$ (1.6 $\mu$s):

$$I(t) = I_{sing}(0) \cdot e^{−t/\tau_s} + I_{trip}(0) \cdot e^{−t/\tau_t} $$

with $I_{sing}(0)$ and $I_{trip}(0)$ being the initial intensities of the emissions from the singlet and triplet state respectively. These intensities depend on the relative population of these states and it is possible to define two components of the pulse shape: one fast component where the emission from the singlet state is dominant and one slow component, where the emission from the singlet state vanishes and that of the triplet state becomes dominant. Investigation of the ratio of these components gives the possibility to discriminate between different kinds of ionizing radiation. In Fig.\ref{fig:12all} averaged pulse shapes for gamma, alpha, and neutrons are shown.

\begin{figure}[h]
\centering
\subfigure[Top: Scatter plot of the background spectrum with $^{222}$Rn admixture showing the gamma  and the alpha bands; Bottom: The corresponding energy background spectrum with $^{222}$Rn inserted into LAr. The first from the left peak corresponging to 2.6 MeV gammas of the $^{228}$Th chain, the second peaks forms 5.5 and 6.0 MeV alphas of the $^{222}$Rn chain , and the last peak is the remaining 7.68 MeV alpha] 
{
    \label{fig:13a}
    \includegraphics[width=0.45\linewidth]{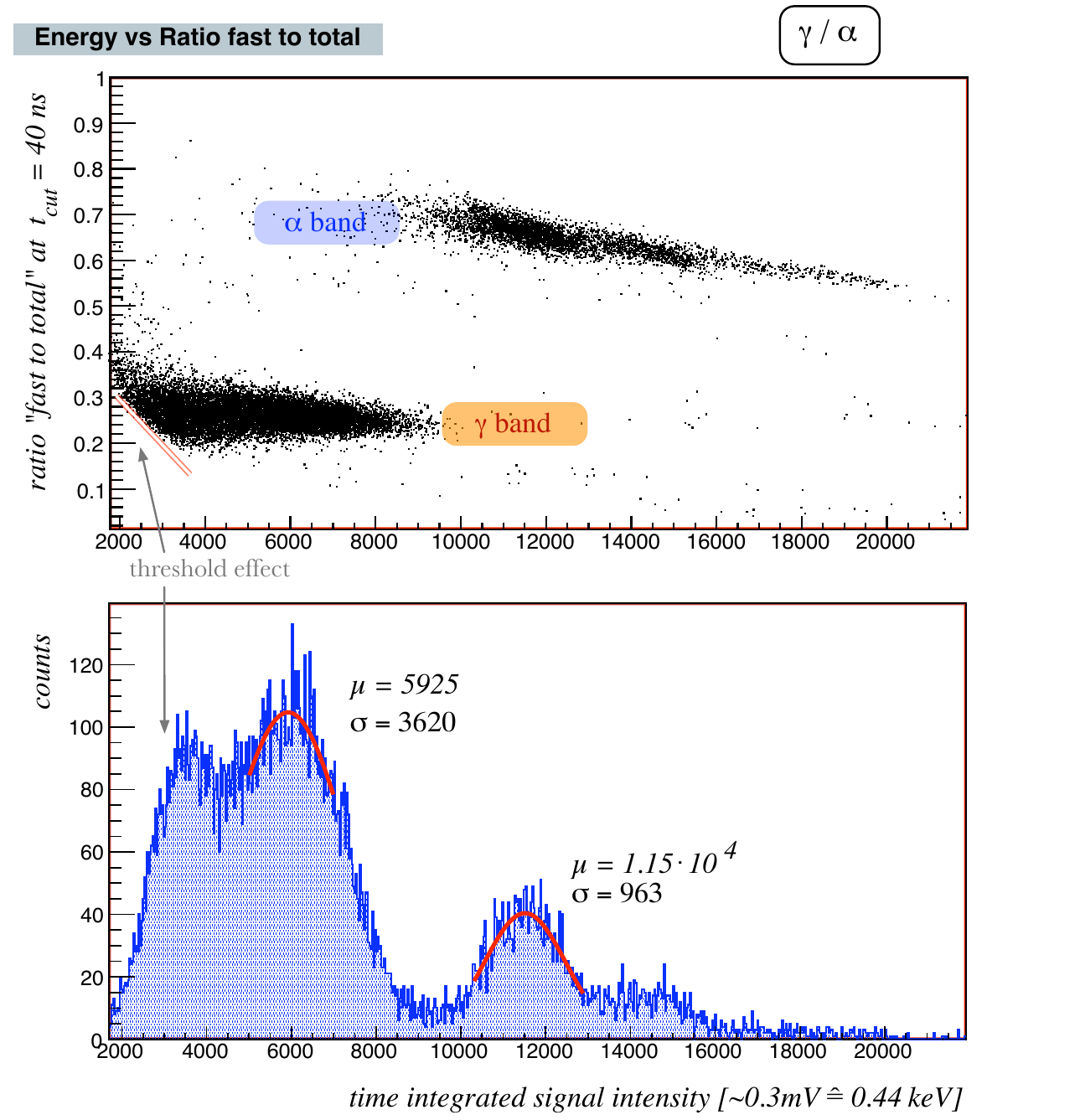}
}
\hspace{0.5cm}
\subfigure[the projection of the scatter plot onto y-axis] 
{
    \label{fig:13b}
    \includegraphics[width=0.45\linewidth]{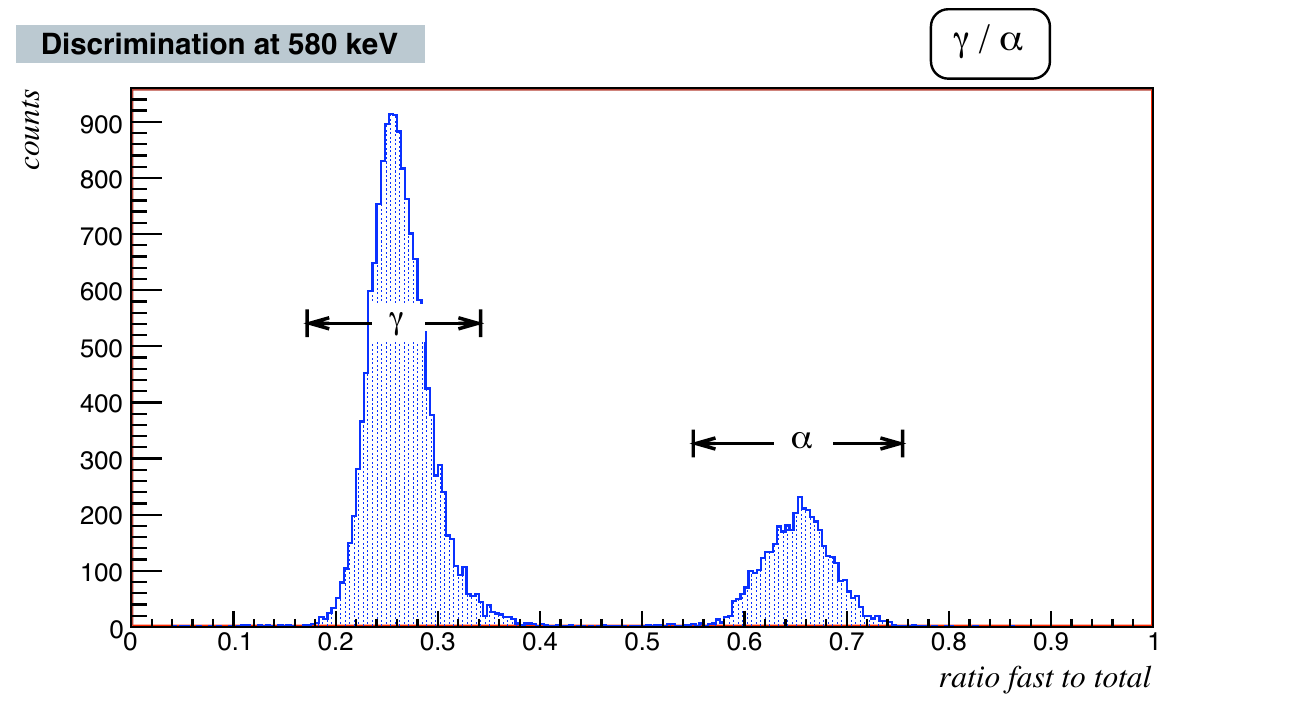}
}
\caption{\label{fig:13all} An example of discrimination between $\alpha$-s and $\gamma$-s.} 
\end{figure}

\begin{figure}[h]
\centering
\subfigure[The scatter plot of the “fast to total” ratio against the energy of the pulses from neutron and gamma radiation] 
{
    \label{fig:14a}
    \includegraphics[height=5cm,width=0.45\linewidth]{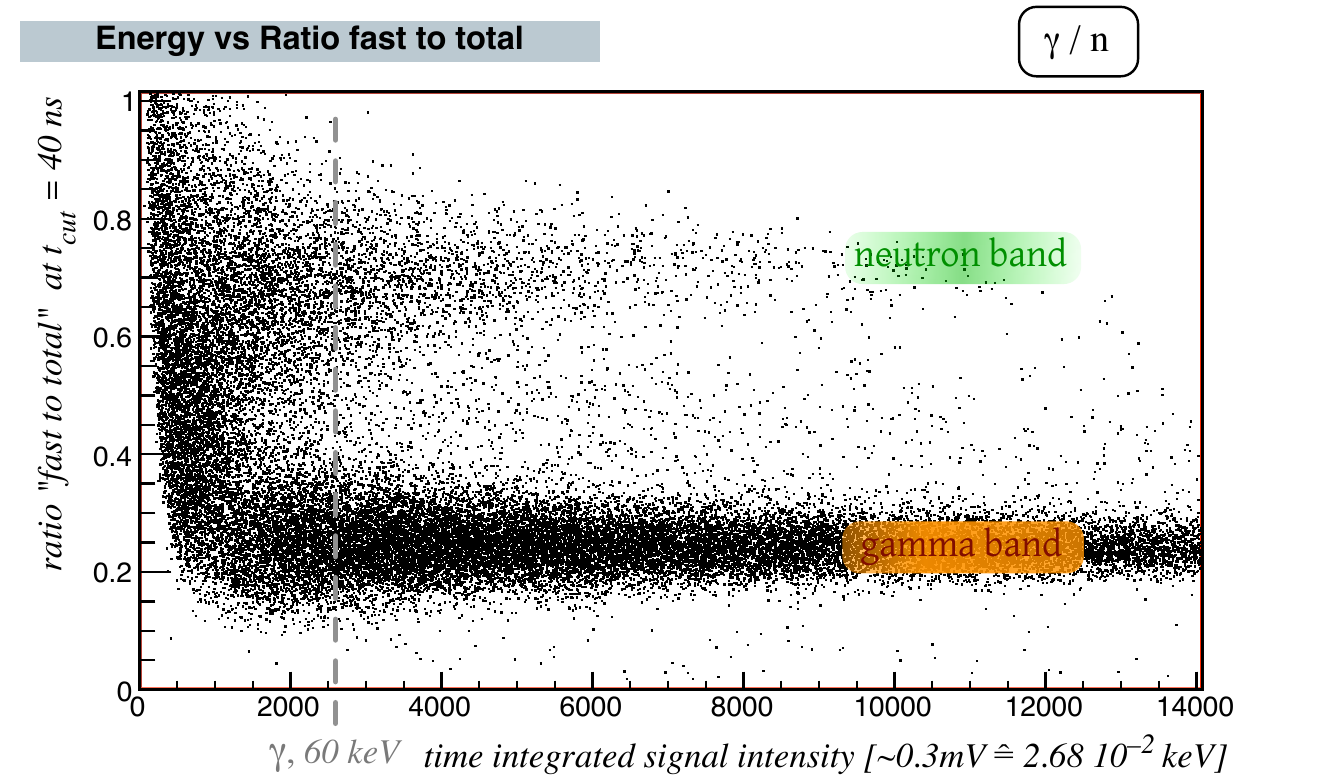}
}
\hspace{0.5cm}
\subfigure[The discrimination power of gamma against neutron signals at an energy threshold of 60 keV. The blue histogram is for the run taken under gamma and neutron radiation. The red histogram shows data from the run with only gamma radiation.] 
{
    \label{fig:14b}
    \includegraphics[width=0.45\linewidth]{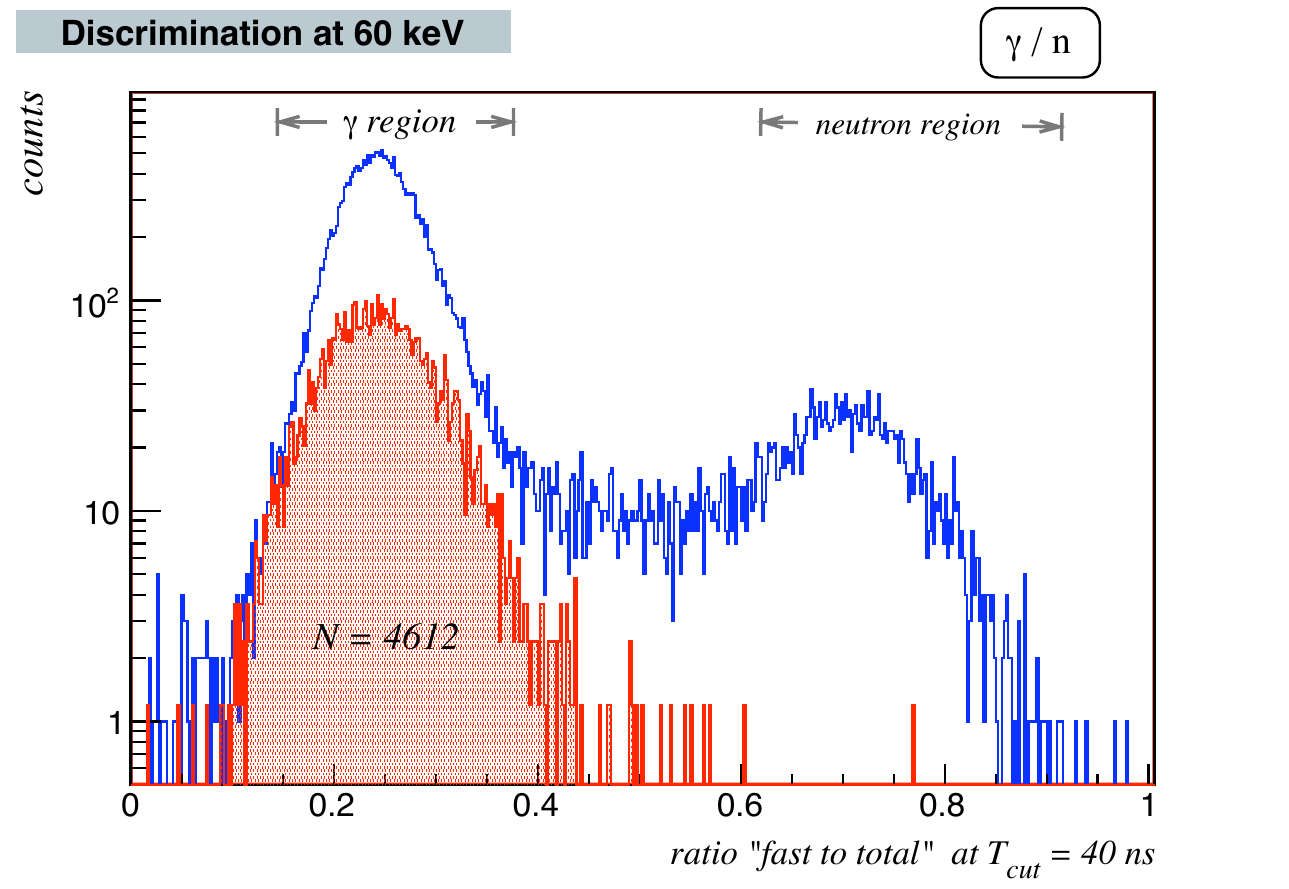}
}
\caption{\label{fig:14all} An example of discrimination between neutrons and $\gamma$-s.} 
\end{figure}

This principle has been successfully used in the mini-LArGe test set up  to discriminate between $\alpha$-s and $\gamma$-s and between $\gamma$-s and neutrons with a probability to misidentify an event of only $P_{mi}(\gamma\alpha) = 5 \cdot 10^{-6}$ and $P_{mi}(\gamma n) < 3 \cdot 10^{-4}$  respectively (see Fig. \ref{fig:13all}, \ref{fig:14all}).

The ability to identify $\alpha$-s can be used in conjunction with the BiPo-trigger technique for a high sensitivity radon level monitoring inside the LAr volume and as a monitor against surface contamination by detecting degraded $\alpha$-s. The ability to reliably detect and identify energy deposition from neutrons in the LAr allows select and eliminate neutron induced background. In particular, this can help to suppress secondary background (like $^{77}$Ge) that is created by the capture of thermal neutrons.

\begin{figure}[h]
\begin{center}
\includegraphics[width=0.7\linewidth]{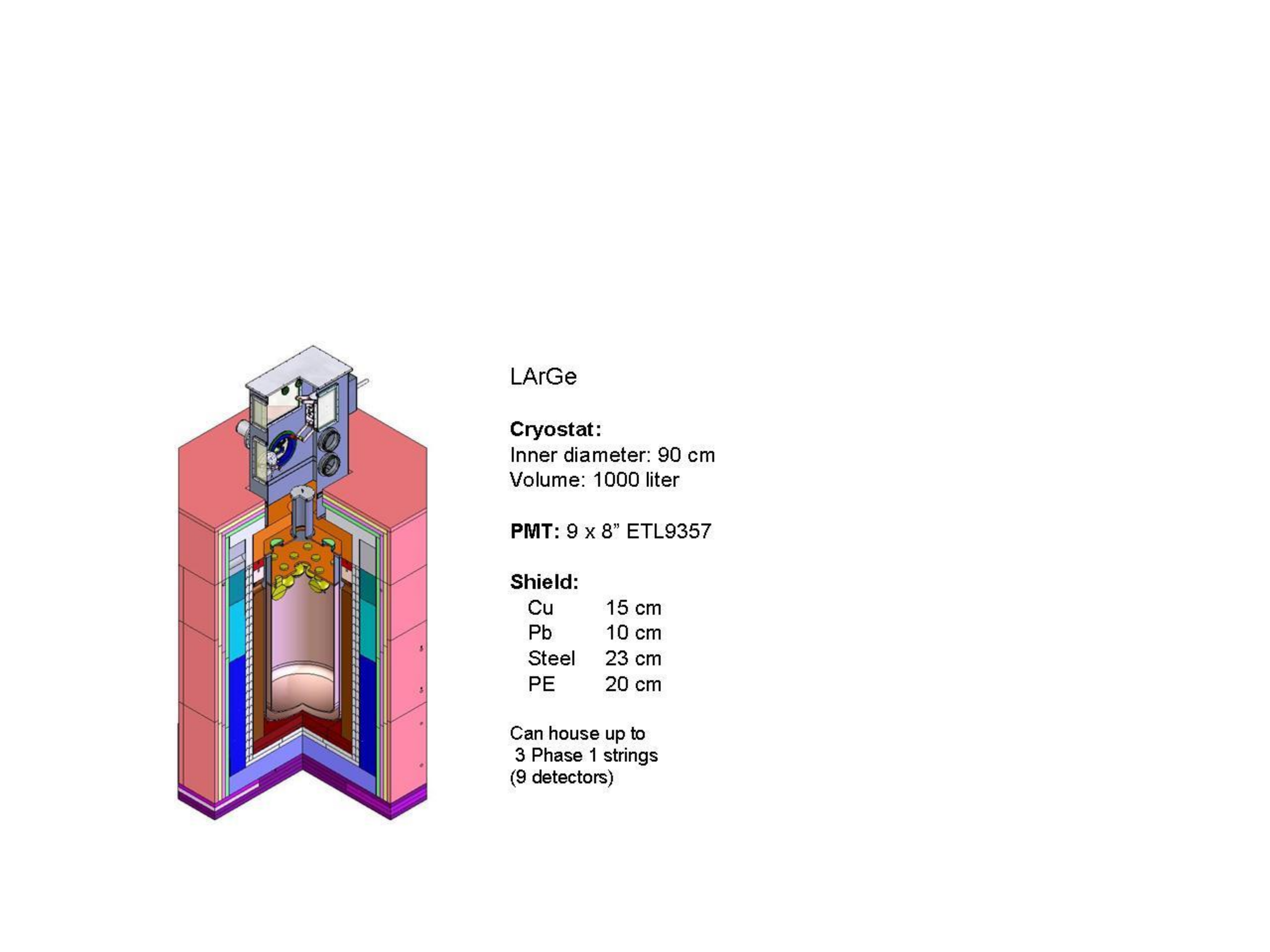}
\caption{\label{fig:15} A schematic drawing of the LArGe setup, which is under construction in the underground facility at LNGS.}
\end{center}
\end{figure}

The results obtained with the mini-LArGe test set up has successfully proven the efficiency of the LAr scintillation as a tool for background suppression and background diagnostics. These investigations will be continued with an active volume of the scale of one cubic meter (1.4 tons) of LAr in the new low background LArGe setup that is currently under construction in the underground facility at LNGS. Figure \ref{fig:15} shows a schematic drawing of the new setup in which a cluster of HP-Ge diodes will be operated in the LAr. 

To detect the scintillation light 9 PMTs (ETL 9357-KFLB) will be used. The high purity copper cryostat will be lined with the improved wavelength shifter and reflector foil developed previously for the mini-LArGe. The setup will be shielded by a multilevel shield consisting of 20 cm poly-ethylene, 23 cm steel, 10 cm lead and 15 cm copper.

\section*{CONCLUSION: Status and outlook}

The \gerda collaboration has submitted the Letter of Intent and the Proposal \cite{gerda} in 2004. \gerda underground operations at LNGS commenced in 2005 with the startup of the new detector laboratory. There, the 8 existing enriched Germanium detectors of HdM and IGEX  are handled, characterized and tested, the same energy resolutions as previously were obtained. Construction of the main experimental infrastructure has started with the order of the water tank and the cryogenic vessel in 2006. The final construction of the water tank will be resumed when the cryostat is erected. The vessel heads of the cryostat have been delivered and the welding is under way. They are being refurbished for mounting in the cryo liquid. 35 kg of new enriched Ge has been procured from Russia for Phase II detectors. Non-enriched and 18-fold segmented detectors have successfully been tested for resolution and pulse shape analysis. Purity levels and purification techniques of liquid argon have been investigated and an extensive program of radioactive purity screening of materials which are used for detector construction is being carried out. New methods of the background suppression such as detector segmentation and anti-coincidence with LAr scintillation have been developed.
The commissioning of the completed setup of GERDA is scheduled for 2008.

The \gerda experiment is partially supported by the DFG within the SFB Transregio 27 "Neutrinos and Beyond", by INTAS (Project Nr. 05-1000008-7996) and by ILIAS-TARI (contract RII-CT-2004-506222). The JINR and INR RAS groups acknowledges the support from RFBR (grants 07-02-01050 and 06-02-16751 respectively).

\clearpage


\end{document}